\documentclass[aps,pre,superscriptaddress,twocolumn,longbibliography]{revtex4}
\usepackage{graphicx,epstopdf,url}
\usepackage[colorlinks=true,urlcolor=blue,citecolor=blue,linkcolor=blue,urlcolor=blue]{hyperref}
\usepackage[english]{babel}
\usepackage[active]{srcltx}
\usepackage{multirow}
\usepackage[version=2]{mhchem}
\begin{document}
\title{
Thermal Stability of Encapsulated Molecular Structures with Extended OH-hydrogen-bond Chains
}
\author{Alexander V. Savin}
\email{asavin@chph.ras.ru}
\affiliation{
N.N. Semenov Federal Research Center for Chemical Physics, Russian Academy of Sciences, Moscow, Russia}
\affiliation{
Plekhanov Russian University of Economics, Moscow, Russia}

\begin{abstract}
Using a coarse-grained model, we performed numerical simulations of the dynamics of molecules forming linear chains on a flat substrate (on the surface of a hexagonal boron nitride (h-BN) crystal).
It is shown that molecules containing benzene rings and hydroxyl groups in their structure can form stable hydrogen-bond chains \ce{OH}$\cdots$\ce{OH}$\cdots$\ce{OH}.
Such chains can be formed by molecules of phenol \ce{C6H5OH}, 4-phenylphenol \ce{C6H5-C6H4OH}, paracetamol \ce{CH3C(O)NHC6H4OH}, and 4-hydroxybenzanilide \ce{C6H5C(O)NHC6H4OH}.
The disruption of the chains for these molecules occurs at temperatures above $T_1=160$, 210, 300, and 390~K, respectively.
Covering such molecular systems with a hexagonal boron nitride sheet (van der Waals encapsulation) significantly enhances their thermal stability.
Such encapsulated molecular structures retain their hydrogen-bond chains up to temperatures $T_2=510$, 860, 600, and 1260~K.
The performed simulations allow us to conclude that chains of these molecules encapsulated by h-BN sheets can be used to create anhydrous proton-exchange membranes capable of operating at high temperatures.
The most promising are chains of 4-hydroxybenzanilide molecules, since their hydrogen-bond chains possess the highest thermal stability.

\end{abstract}

\maketitle
\section{Introduction}
The presence in a molecular system of chains of hydroxyl groups linked by hydrogen bonds,
\begin{equation}
\ce{O-H}\cdots\ce{O-H}\cdots\ce{O-H}\cdots\ce{O-H}\cdots\ce{O-H}\cdots
\label{f1}
\end{equation}
ensures high proton conductivity in the direction of these chains~\cite{Zundel2000}.
Proton transport across cell membranes occurs through protein proton channels and proceeds along hydrogen-bond chains~\eqref{f1} formed by amino acid residues containing a hydroxyl group \ce{OH} (serine, threonine, tyrosine)~\cite{Nagle1978,Kaliman2008,Paulino2020}.
In the bacteriorhodopsin molecule, the hydrogen-bond chain is formed by tyrosine residues contained in the seven transmembrane $\alpha$-helical segments of the molecule~\cite{Merz1981}.
Hydrogen-bond chains~\eqref{f1} act as proton conductors, providing an efficient pathway for fast proton transfer~\cite{Fillaux2002}.

The idea of proton transfer along hydrogen-bond chains was proposed by Theodor von Grotthuss as early as 1804~\cite{Grotthuss1806,Marx2006,Cukierman2006}.
According to modern concepts, proton transfer in water and ice occurs along the hydrogen-bond network and is divided into two stages: the passage of an ionic defect (\ce{H+}) and the passage of an orientational defect (Bjerrum defect), which, after the passage of the ionic defect, returns the hydrogen-bond chain to its original state~\cite{Bjerrum1952,Nagle1978,Merz1981}.
The highest proton conductivity is possessed by phosphoric acid \ce{H3PO4}, which is capable of forming branched hydrogen-bond chains~\cite{Vilciauskas2012}.

At present, the creation of molecular systems with high proton conductivity is an important task for the development and improvement of proton-exchange membranes (PEMs), whose main function is to conduct protons (\ce{H+}) while being an insulator for electrons and a barrier for gases~\cite{Kiani2025,Luo2026}.
PEM fuel cells have demonstrated high potential as environmentally clean alternative energy sources.
However, their use is hindered by the degradation of PEM performance at high temperatures and low humidity.
To solve this problem, it is necessary to create new materials with high proton conductivity at elevated temperatures and in the absence of water.
Such materials must consist of molecules capable of creating hydrogen-bond chains~\eqref{f1} that are stable at high temperatures.

\begin{figure}[tb]
\begin{center}
\includegraphics[angle=0, width=1.0\linewidth]{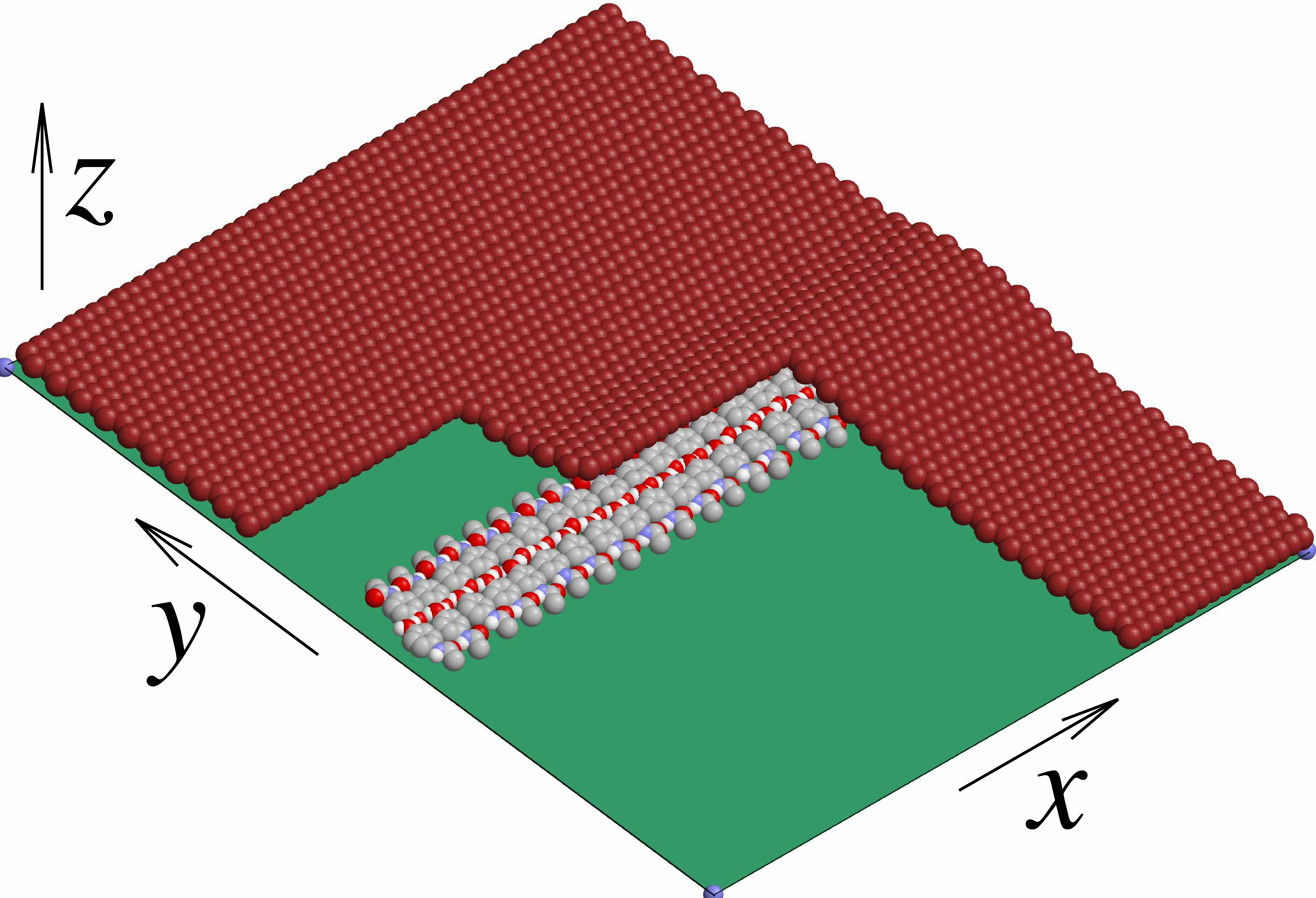}
\end{center}
\caption{\label{fig01}\protect
Linear structure of 46 paracetamol molecules located on a flat substrate $z\le 0$ and covered from above by an h-BN sheet.
The calculated periodic cell of size $13.02\times 13.01$~nm$^2$ at $T=400$~K is shown;
the substrate surface is shown in green, hydrogen atoms in white, carbon in gray, nitrogen in light blue, and oxygen in red.
The paracetamol molecules form three parallel hydrogen-bond chains;
the chain of hydroxyl groups \ce{OH}$\cdots$\ce{OH}$\cdots$\ce{OH} is located in the center (red-white line).
The atoms of the covering sheet are shown as brown spheres
(for clarity, only part of the sheet atoms is shown).
  }
\end{figure}

In this work, using the molecular dynamics method and coarse-grained models, we analyze the possibility of creating such materials from planar molecules located on a flat substrate and covered from above (encapsulated) by a hexagonal boron nitride (h-BN) sheet.
We will consider encapsulated chains of molecules of phenol \ce{C6H5OH},
4-phenylphenol \ce{C6H5-C6H4OH}, paracetamol \ce{CH3C(O)NHC6H4OH},
and 4-hydroxybenzanilide \ce{C6H5C(O)NHC6H4OH} --- see Fig.~\ref{fig01}.
The structures of these molecules contain benzene rings, which ensure their strong interaction with the flat substrate, and hydroxyl groups \ce{OH}, which allow them to form extended hydrogen-bond chains~\eqref{f1} necessary for proton transport.
Hexagonal boron nitride, unlike graphene, is an insulator with extremely low electrical conductivity,
and therefore it is better suited for creating PEMs.
For this reason, the flat surface of an h-BN crystal will be considered as the substrate.

Covering a molecular system located on a flat substrate with an h-BN sheet (van der Waals encapsulation)
creates a trap effect for it.
Due to the van der Waals interaction of the sheet with the flat substrate, an internal pressure of the order of several GPa can arise in such a nanoscale pocket~\cite{Vasu2016,Khestanova2016,Hu2024}, which can substantially change the properties of the enclosed material.
Van der Waals encapsulation is a convenient method for creating high local pressure to modify the properties of a system~\cite{Zhang2017} or for its storage~\cite{Zheng2012,Slepchenkov2018,Apkadirova2022}.
Encapsulation of molecular systems with hydrogen-bond chains should substantially enhance their stability.

\section{Model}
\begin{figure}[tb]
\begin{center}
\includegraphics[angle=0, width=1.0\linewidth]{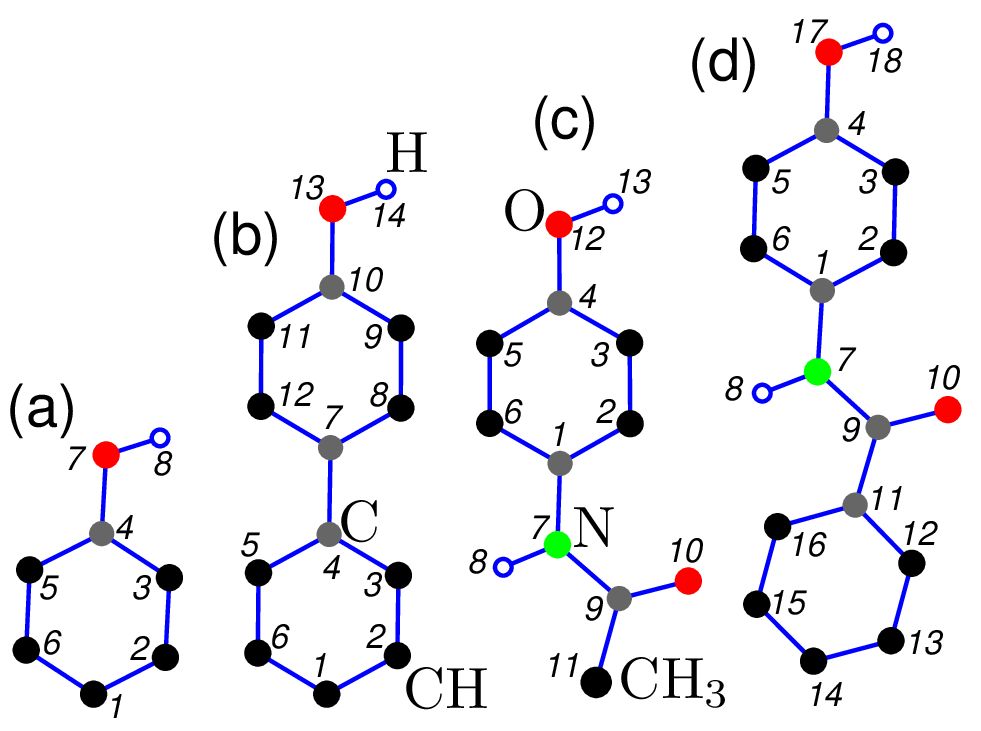}
\end{center}
\caption{\label{fig02}\protect
Coarse-grained models for molecules of (a) phenol \ce{C6H5OH}, (b) 4-phenylphenol \ce{C6H5-C6H4OH}, (c) paracetamol \ce{CH3C(O)NHC6H4OH}, and (d) 4-hydroxybenzanilide \ce{C6H5C(O)NHC6H4OH}.
 Hydrogen atoms are shown as white spheres, carbon as gray, nitrogen as green,
oxygen as red, and the united atoms \ce{CH} and \ce{CH3} as black (the atom numbering used for the molecules is shown). Blue lines indicate valence bonds.
  }
\end{figure}

When simulating the dynamics of molecules of phenol \ce{C6H5OH} (PhOH), 4-phenylphenol \ce{C6H5-C6H4OH} (4-PP), paracetamol \ce{CH3C(O)NHC6H4OH} (PCM), and 4-hydroxybenzanilide \ce{C6H5C(O)NHC6H4OH} (4-HBA), we use the united-atom approximation, in which the molecular groups CH and CH$_3$ are treated as united atoms whose centers coincide with the centers of the carbon atoms --- see Fig.~\ref{fig02}.
In this approximation, the PhOH molecule consists of $N_0=8$, the 4-PP molecule of $N_0=14$, PCM of $N_0=13$, and the 4-HBA molecule of $N_0=18$ united atoms.
The masses of the united atoms are given in Table~\ref{tab1}.
\begin{table*}[tb]
\caption{
Masses and interaction potential parameters for united atoms X:
$M_i$ --- atom mass ($m_p=1.6603\times 10^{-27}$~kg --- proton mass),
$\varepsilon_i$ and $r_i$ --- energy and radius of the LJ interaction,
$q_i$ --- electric charge of the atom,
$\epsilon_i$ and $h_i$ --- interaction energy and equilibrium distance to the flat substrate
(h-BN crystal surface), $i$ --- atom number in the molecule (the numbering for the paracetamol molecule is given).
\label{tab1}
}
\begin{center}
\setlength{\tabcolsep}{7pt}
\begin{tabular*}{0.85\linewidth}{ccccccccccc}
\hline\hline
 X                 &\ce{C}&\ce{C}&\ce{CH}   &\ce{N}& \ce{H}&\ce{C}&\ce{O}&\ce{CH3}&\ce{O}&\ce{H}\\
 $i$               &   1  & 4    &2, 3, 5, 6&  7   & 8     & 9    & 10   & 11     & 12   & 13\\

\hline
 $M_i$~($m_p$)     &  12  & 12   & 13       & 14   & 1     & 12   & 16   & 15     & 16   &  1\\
$\varepsilon_i$~(meV)&4.284&4.284&4.284     &4.080 &0.434  &4.284 &6.344 &4.284   &6.344 &0.434\\
$r_i$~(\AA)        &1.861 &1.861 &1.861     &1.899 &0.621  &1.861 &1.711 &1.861   &1.711 &0.621\\
$q_i$~(e)          &0.066 &0.100 & 0        &-0.463&0.286  &0.580 &-0.504&0.035   &-0.500&0.400\\
$\epsilon_i$~(meV) &83.5  &83.5  &106.7     &116.9 &11.1   & 83.5 & 92.7 &123.8   & 79.0 &5.7\\
$h_i$~(\AA)        &3.34  &3.34  &3.29      &3.28  &2.25   & 3.34 & 3.21 & 3.77   & 3.31 &1.97\\
\hline\hline
\end{tabular*}
\end{center}
\end{table*}

We use a force field in which various potentials describe the deformations of valence bonds, valence and dihedral (torsion) angles, as well as nonvalent interactions of atoms.
In this field, the deformation energy of the valence bonds \ce{C-CH}, \ce{CH-CH}, \ce{C-N}, \ce{N-H}, \ce{C=O}, \ce{C-CH3}, \ce{C-O}, and \ce{O-H} is described by the harmonic potential
\begin{equation}
U_{\rm v}(\rho)=\frac12K(\rho-\rho_0)^2,
\label{f2}
\end{equation}
where $\rho$ and $\rho_0$ are the current and equilibrium bond lengths, and $K$ is the bond stiffness.
The values of the parameters of potential \eqref{f2} for various valence bonds are given in Table~\ref{tab2}.
\begin{table}[tb]
\caption{
Values of the parameters of the harmonic potential \eqref{f2} for various valence bonds X--Y.
\label{tab2}
}
\begin{center}
\setlength{\tabcolsep}{2pt}
\begin{tabular*}{\linewidth}{cccccccc}
\hline\hline
 \ce{X-Y}      &\ce{C-CH}      &\ce{C-N}&\ce{N-H}&\ce{C=O}&\ce{C-CH3}&\ce{C-O}&\ce{O-H}\\
  ~            & \ce{CH-CH}    & ~      &   ~    &  ~     & ~        & ~      & ~ \\
\hline
 $K$~(N/m)     & 469           & 427    & 434    & 570    & 553      & 450    & 317\\
 $\rho_0$~(\AA)& 1.39          &1.405   & 1.007  &1.222   & 1.505    & 1.364  & 0.96\\
\hline\hline
\end{tabular*}
\end{center}
\end{table}

The deformation energy of valence angles \ce{X-Y-Z} is described by the potential
\begin{equation}
U_a({\bf u}_1,{\bf u}_2,{\bf u}_3)=\varepsilon_a(\cos\varphi-\cos\varphi_0)^2,
\label{f3}
\end{equation}
where the cosine of the valence angle is $\cos\varphi=-({\bf v}_1,{\bf v}_2)/\rho_1\rho_2$,
the vectors ${\bf v}_1={\bf u}_2-{\bf u}_1$, ${\bf v}_2={\bf u}_3-{\bf u}_2$,
and the bond lengths $\rho_1=|{\bf v}_1|$, $\rho_2=|{\bf v}_2|$.
Here the vectors ${\bf u}_1$, ${\bf u}_2$, and ${\bf u}_3$ specify the coordinates of the atoms forming the valence angle $\varphi$, and $\varphi_0$ is the equilibrium value of the angle.
The values of the potential parameters for various valence angles are given in Table~\ref{tab3}.

The deformation of a dihedral angle is described by the potential
\begin{equation}
U_d({\bf u}_1,{\bf u}_2,{\bf u}_3,{\bf u}_4)=\epsilon_d(1+z_d\cos\phi),
\label{f4}
\end{equation}
where $\cos\phi=({\bf w}_1,{\bf w}_2)/|{\bf w}_1||{\bf w}_2|$,
the vectors ${\bf w}_1=({\bf u}_2-{\bf u}_1)\times ({\bf u}_3-{\bf u}_2)$ and
${\bf w}_2=({\bf u}_3-{\bf u}_2)\times ({\bf u}_4-{\bf u}_3)$.
The values of the potential parameters used for various dihedral angles are given in Table~\ref{tab4}.

For a pair of atoms X$_i$, X$_j$ ($i$, $j$ are atom numbers in the molecule) participating in the formation of the dihedral angle X$_i$--Y--Z--X$_j$, their nonvalent interaction is also taken into account, described by the Lennard-Jones (LJ) potential
\begin{equation}
U_{LJ}(r)=\varepsilon_0[(r_0/r)^{12}-2(r_0/r)^6],
\label{f5}
\end{equation}
with interaction energy $\varepsilon_0=\sqrt{\varepsilon_i\varepsilon_j}/2$, where
$r$ is the current distance between the interacting atoms and the equilibrium distance is $r_0=r_i+r_j$.
For paracetamol and 4-hydroxybenzanilide molecules, the LJ interaction of the oxygen atom of the amide group with the united CH atoms ($i=2$, 6 and $i=2$, 6, 12, 16) is also taken into account,
with interaction energy $\varepsilon_0=\sqrt{\varepsilon_i\varepsilon_{10}}$ and equilibrium distance $r_0=r_i+r_{10}$.
The values of the parameters $\varepsilon_i$ and $r_i$ are given in Table~\ref{tab1}.

For the 4-PP molecule, the valence bond connecting the benzene rings [see Fig.~\ref{fig02}~(b)] is a single bond, and the planes of adjacent rings in isolated molecules form a dihedral angle $\phi_t=134^\circ$.
Here, to describe the deformation of the dihedral angles formed by the atoms C$_3$C$_4$C$_7$C$_{12}$ and C$_5$C$_4$C$_7$C$_8$, one must use the potential
\begin{equation}
U_t({\bf u}_1,{\bf u}_2,{\bf u}_3,{\bf u}_4)=\epsilon_t(\cos\phi-\cos\phi_t)^2,
\label{f6}
\end{equation}
where the energy is $\epsilon_t=0.045$~eV.
\begin{table}[tb]
\caption{
Values of the parameters for the valence-angle potential \ce{X-Y-Z} \eqref{f3}.
\label{tab3}
}
\begin{center}
\setlength{\tabcolsep}{1.5pt}
\begin{tabular*}{\linewidth}{cccccccccc}
\hline\hline
 XYZ          & CCC & CCN & CNH & CNC & NCO & NCC & OCC & CCO & COH\\
\hline
$\varepsilon_a$~(eV)   & 3.643&3.823    & 2.781  & 4.888 &4.932   & 3.758  & 4.625  & 4.047  & 1.791\\
$\varphi_0$~($^\circ$)&120 & 117     &  118   & 128   & 123    & 116    & 120    & 120    & 113\\
\hline\hline
\end{tabular*}
\end{center}
\end{table}
\begin{table*}[tb]
\caption{
Values of the parameters for the dihedral-angle potential \ce{X-Y-Z-W} \eqref{f4} for different atoms
(the atom numbering for the paracetamol molecule is used).
\label{tab4}
}
\begin{center}
\setlength{\tabcolsep}{5pt}
\begin{tabular*}{0.82\linewidth}{cccccccccc}
\hline\hline
 XYZW            & CCCC & CCCN & C$_6$C$_1$NH & C$_2$C$_1$NC & C$_6$C$_1$NC &C$_2$C$_1$NC & CNCO & CNCC$_{11}$ & CCCO\\
\hline
$\epsilon_d$~(eV)& 0.63 & 0.63 & 0.42   & 0.42   & 0.42 & 0.42    & 0.42  & 0.42  & 0.63\\
$z_d$ & -1   & 1    &  -1    &  1     &  1   & -1      & -1    &   1   & 1\\
\hline\hline
\end{tabular*}
\end{center}
\end{table*}
\begin{table*}[tb]
\caption{
Values of the parameters for the LJ potential \eqref{f5} for the interaction of united atoms X
with a united atom of the h-BN sheet.
\label{tab5}
}
\begin{center}
\begin{tabular*}{0.71\linewidth}{ccccccccc}
\hline\hline
 X           & \ce{C}  &\ce{CH}&\ce{CH3}&\ce{N}&~~\ce{O}~(C=O)~~&~~\ce{O}~(OH)~~&~~\ce{H}~(NH)~~&~~\ce{H}~(OH)~~\\
\hline
$\varepsilon_i$~(meV)&~~~~9.37~~&~~11.99~~&~~13.53~~& ~13.67~& ~11.40~      & ~9.09~      & ~2.98~ & ~2.04~\\
$r_i$~(\AA)      & ~~3.77    & 3.72    & 4.19    & 3.71   & 3.62         &3.73       & 2.53      & 2.22\\
\hline\hline
\end{tabular*}
\end{center}
\end{table*}

The interaction between molecules is described by the potential
\begin{equation}
U({\bf X}_1,{\bf X}_2)\!=\!\sum_{i=1}^{N_0}\sum_{j=1}^{N_0}\!\left\{\varepsilon_{ij}\!
\left[\!\left(\frac{\bar{r}_{ij}}{r_{ij}}\right)^{12}\!-\!2\!\left(\frac{\bar{r}_{ij}}{r_{ij}}\right)^{6}\!\right]\!+\!\kappa \frac{q_iq_j}{r_{ij}}\!\right\},
\label{f7}
\end{equation}
where $N_0$ is the number of united atoms in the molecule, and the $3N_0$-dimensional vector
${\bf X}_k=\{{\bf u}_{k,i}\}_{i=1}^{N_0}$ ($k=1,2$) specifies the coordinates of the molecule atoms (the vector ${\bf u}_{k,i}$ specifies the position of the $i$-th atom of the $k$-th molecule), and the distance between atoms is $r_{ij}=|{\bf u}_{1,i}-{\bf u}_{2,j}|$.
Here the energy is $\varepsilon_{ij}=\sqrt{\varepsilon_i\varepsilon_j}$, the equilibrium distance is
$\bar{r}_{ij}=r_i+r_j$, and $q_i$ is the electric charge of the atom ($i,j=1,\dots,N_0$), with the coefficient
$\kappa=14.400611$~eV\AA/e$^2$.
The values of the parameters $\varepsilon_i$, $r_i$, and $q_i$ are given in Table~\ref{tab1}.
All parameter values of the interaction potentials \eqref{f2}--\eqref{f7} were obtained from
the AMBER General Force Field (version 2.1, April 2016)~\cite{Amber}.

For the substrate, we use the approximation of a fixed attractive plane.
In this approximation, the van der Waals interaction of the molecule atoms with the flat substrate can be described by the LJ potential $(m,l)$
\begin{equation}
W({\bf X})\!=\!\sum_{i=1}^{N_0}\!W_i(z_i)\!=\!\sum_{i=1}^{N_0}\frac{\epsilon_i}{l-m}
\left[ m\left(\frac{h_i}{z_i}\right)^l-l\left(\frac{h_i}{z_i}\right)^m\right],
\label{f8}
\end{equation}
where $z_i$ is the distance from the $i$-th atom to the outer surface of the flat substrate $z\le 0$.
The potential $W_i(z_i)$ specifies the dependence of the interaction energy of the $i$-th atom on its distance to the substrate.
This dependence was obtained numerically for various substrates~\cite{Savin2019,Savin2021}.
The potential $W_i(z_i)$ has a minimum $W_i(h_i)=-\epsilon_i$ ($\epsilon_i$ is the binding energy of the atom to the substrate).
As the flat substrate, we use the surface of an h-BN crystal, for which
the exponents are $l=10$, $m=3.75$.
The values of the parameters $\{\epsilon_i, h_i\}_{i=1}^{N_0}$ are given in Table~\ref{tab1}.
Note that an h-BN sheet, unlike a graphene sheet, is not a conductor.
Therefore, when constructing structures for proton-exchange membranes, it is more preferable.

\begin{figure}[tb]
\begin{center}
\includegraphics[angle=0, width=1.0\linewidth]{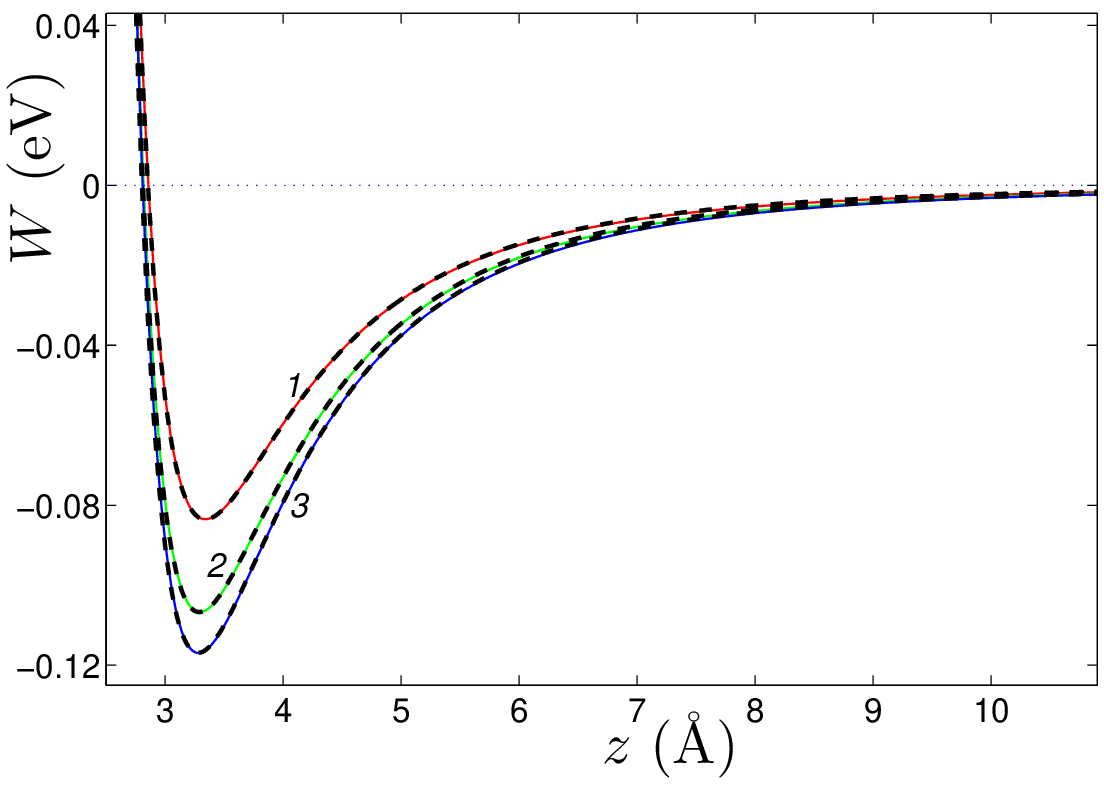}
\end{center}
\caption{\label{fig03}\protect
Dependence of the interaction energy $W(z)$ of C, CH, and N atoms (curves 1, 2, and 3) with the flat surface $(z=0)$ of an h-BN crystal on the distance to the surface $z$.
Solid curves give the dependences obtained numerically.
Dashed lines correspond to the LJ potential (\ref{f10}) with the parameter values $\epsilon_0$ and $h_0$ from Table~\ref{tab1}.
}
\end{figure}

The interaction of a particle with the half-space $z\le 0$ is usually described by the LJ potential (3,9)
\begin{equation}
W(z)=\epsilon_0[(h_0/z)^9-3(h_0/z)^3]/2, \label{f9}
\end{equation}
where $\epsilon_0$ is the interaction energy (adhesion energy), and $h_0$ is the equilibrium distance to the surface of the half-space \cite{Aitken2010,Zhang2013,Zhang2014}.
The interaction of a particle with the surface of an h-BN crystal is more accurately described by the LJ potential
\begin{equation}
W(z)=\frac{\epsilon_0}{(l-m)}[m(h_0/z)^l-l(h_0/z)^m], \label{f10}
\end{equation}
with exponents $l=10$, $m=3.75$ --- see Fig.~\ref{fig03}.
This difference is due to the fact that here the half-space has a layered structure and the main contribution comes from the interaction with the first layer (with the upper sheet of the crystal).

The potential \eqref{f10} was found numerically as the dependence of the interaction energy of a particle
(a united atom) with a 10-layer h-BN sheet on the distance $z$ to the sheet surface.
The calculations used sheets of size $5.06\times 4.89$~nm$^2$ consisting of 960 atoms.
The surface of the upper sheet coincided with the plane $z=0$, and the particle was located above its center
at a distance $z>0$.
The interaction energy $W(z)$ was calculated as the sum of all pairwise interactions between the particle and the atoms of the multilayer sheet using the LJ potentials \eqref{f5} with parameter values from the AMBER force field \cite{Amber} (the parameter set from version GAFF2 was used).
For the atoms of the h-BN sheet, the parameters
$\varepsilon_{\rm B}=4.119$~meV, $r_{\rm B}=1.938$~\AA~ and
$\varepsilon_{\rm N}=6.281$~meV, $r_{\rm N}=1.889$~\AA~ \cite{Cheng2007} were used.
The dependence $W(z)$ was averaged over the unit cell of the sheet (over the coordinates $x$, $y$ within this cell).
As can be seen from Fig.~\ref{fig03}, the numerically obtained dependence $W(z)$ is described with high accuracy by the potential \eqref{f10}.

\begin{figure}[tb]
\begin{center}
\includegraphics[angle=0, width=1.0\linewidth]{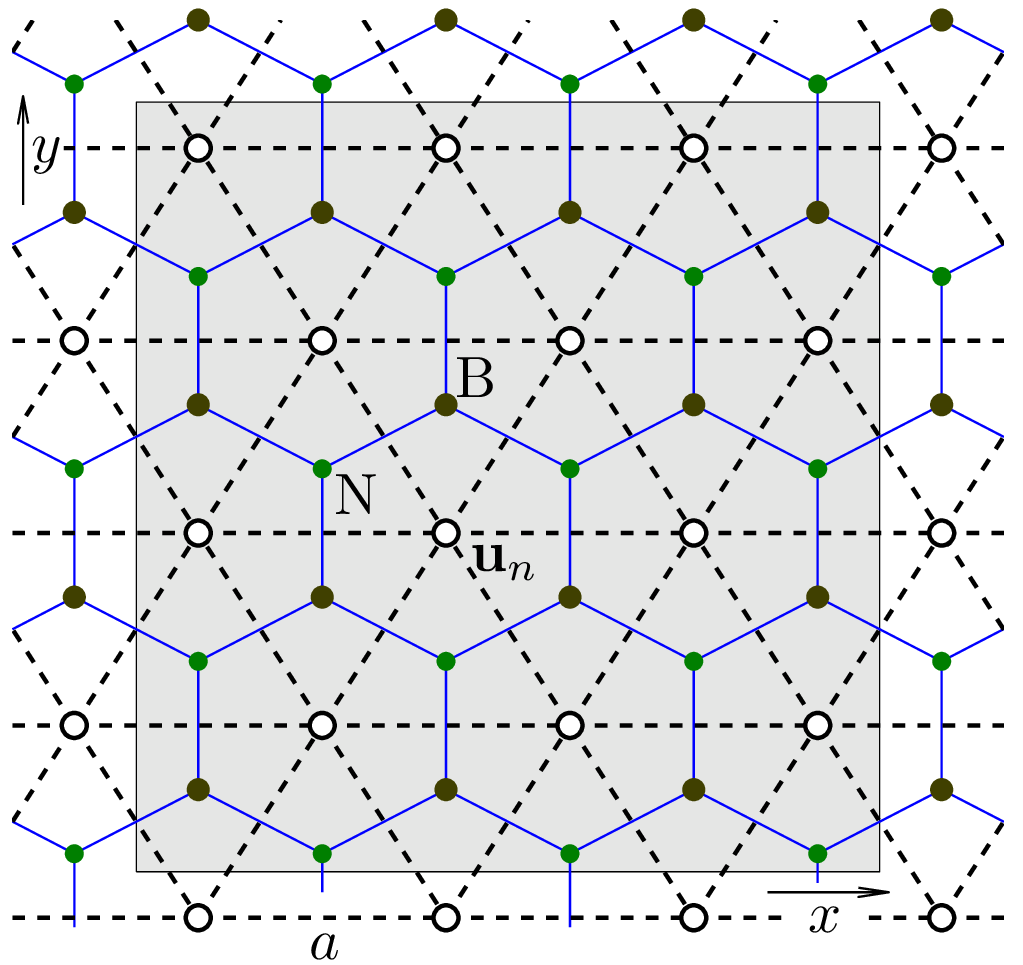}
\end{center}
\caption{\label{fig04}\protect
Construction of the coarse-grained model for an h-BN sheet.
The hexagonal lattice of the sheet is shown (nitrogen atoms correspond to green spheres,
boron to brown spheres; blue lines show the \ce{B-N} valence bonds).
Large white spheres show the positions of the united atoms, and dashed lines show the bonds between them.
The gray square shows the periodic cell of the triangular lattice of size $3a\times 4a_y$;
the lattice spacing is $a=\sqrt{3}r_{\rm BN}$, $a_y=\sqrt{3/4}a$.
}
\end{figure}

To simulate the h-BN sheet covering the molecular system from above, we also use a coarse-grained model in which a pair of neighboring B--N atoms corresponds to one united atom of mass $M_0=M_{\rm B}+M_{\rm N}=24.82m_p$ -- see Fig.~\ref{fig04}.
The united atoms are located at the centers of the hexagons of the valence bonds of the sheet.
The lines between neighboring united atoms form a lattice of equilateral triangles that triangulate the sheet.
Each triangle corresponds to one atom of the sheet located at its center.

It is convenient to regard the sides of the triangles as ``valence'' bonds between neighboring united atoms.
We describe them by the harmonic potential \eqref{f2} with stiffness coefficient $K_0=270$~N/m and equilibrium distance $\rho_0=a=2.5045$~\AA~ (the triangular lattice spacing $a=\sqrt{3}r_{\rm BN}$, the valence bond length $r_{\rm BN}=1.446$~\AA).
With these values, the elastic and geometric properties of the triangular lattice best correspond to the properties of a flat h-BN sheet.
The sound velocity for the longitudinal acoustic mode of the lattice is $v_{LA}=a\sqrt{9K_0/8M_0}=21.50$~km/s,
and for the transverse mode $v_{TA}=a\sqrt{3K_0/8M_0}=12.41$~km/s (for an h-BN sheet the velocities are
$v_{LA}=21$~km/s, $v_{TA}=12.71$~km/s~\cite{Michel2009}).

We take into account the bending deformations of the triangular lattice using the dihedral-angle potential between adjacent triangles \eqref{f4}, where the sign is $z_d=1$ and the energy $\epsilon_d>0$.
In this case, each ``valence'' bond between united atoms corresponds to one dihedral angle with equilibrium value $\phi_0=\pi$.

The value of the energy $\epsilon_d$ can be determined from the bending stiffness of a single sheet $\kappa$.
To this end, consider an h-BN nanotube oriented along the $x$ axis, whose cross section consists of $N$ segments.
The radius of such a nanotube is $R=\frac12 a_y/\sin(\pi/N)$ ($a_y=a\sqrt{3/4}$), and the energy of a cylindrical fragment of length $a$ is
$E=2N\epsilon_d\sin^2(\pi/N)$, with its surface area $S=a^2N\sqrt{3/4}$.
Since in the limit $N\rightarrow\infty$ the energy is $E=\frac12\kappa S/R^2$, we have
$\kappa=2ER^2/S=\epsilon_d\sqrt{3/4}$. Therefore, the energy is $\epsilon_d=2\kappa/\sqrt{3}$.
According to first-principles calculations, the bending stiffness of an h-BN sheet is $\kappa=1.563$~eV \cite{Sgouros2025}, so the energy is $\epsilon_d=1.80$~eV.

The interaction of the united atoms of the sheet with the flat substrate is also described by the LJ potential \eqref{f10}
with exponents $l=10$, $m=3.75$, energy $\epsilon_0=0.1883$~eV, and equilibrium distance $h_0=3.39$~\AA.

The interaction of the united atoms of the covering sheet with the atoms of the molecules is described with good accuracy by the LJ potentials \eqref{f5} with the parameter values $\epsilon_0$ and $r_0$ given in Table~\ref{tab5}.
To obtain these values, the dependence of the energy of the nonvalent interaction of a molecule atom with a pair of B and N atoms located at a distance $r_{\rm BN}$ was calculated.

A similar coarse-grained model can also be constructed for a graphene sheet.
Coarse-grained models of graphene are characterized by the number of atoms $N_{\rm CG}$ corresponding to one united atom.
Models with $N_{\rm CG}=4$~\cite{Ruiz2015,Shang2017,Liu2021,Wang2023},
$N_{\rm CG}=6$, 8, ... \cite{Kauzlaric2011,Wang2013}, having a hexagonal lattice structure, are usually used.
Such models lead to lattices with very large cells and are therefore inconvenient for simulating van der Waals encapsulations (molecular systems covered by a graphene sheet).
For the proposed h-BN sheet model, the coarse-graining number is $N_{\rm CG}=2$, and the small size of the triangular cells allows its use in simulating encapsulations.
The simplicity of the model (triangular lattice structure and the need to use only two potentials for each bond) makes it possible to significantly speed up the numerical simulations.

\section{Encapsulated molecular chains}

Consider a two-component molecular system formed by an h-BN sheet lying on a flat substrate,
with molecules located between them (see Fig.~\ref{fig01}).
Let the sheet consist of $N_2$ united atoms, and the molecular component consist of $N_1$ molecules consisting of $N_0$ atoms.
As the substrate, we take the attractive plane $z=0$, corresponding to the flat surface of an h-BN crystal.

Take a rectangular sheet in which some of the bonds between united atoms form lines parallel to the $x$ axis (the ``zigzag'' direction\ of the h-BN sheet coincides with the $x$ axis --- see Fig.~\ref{fig04}).
Let each horizontal line consist of $N_x$ united atoms,
and the sheet itself contain $N_y$ such lines.
Then the rectangular sheet will consist of $N_2=N_xN_y$ united atoms.
In the simulations, we use periodic boundary conditions along the $x$ and $y$ axes with
periods $L_x=N_xa$, $L_y=N_ya_y$, $a_y=a\sqrt{3/4}$, coinciding with the dimensions of this rectangular sheet.
\begin{figure}[tb]
\begin{center}
\includegraphics[angle=0, width=1.0\linewidth]{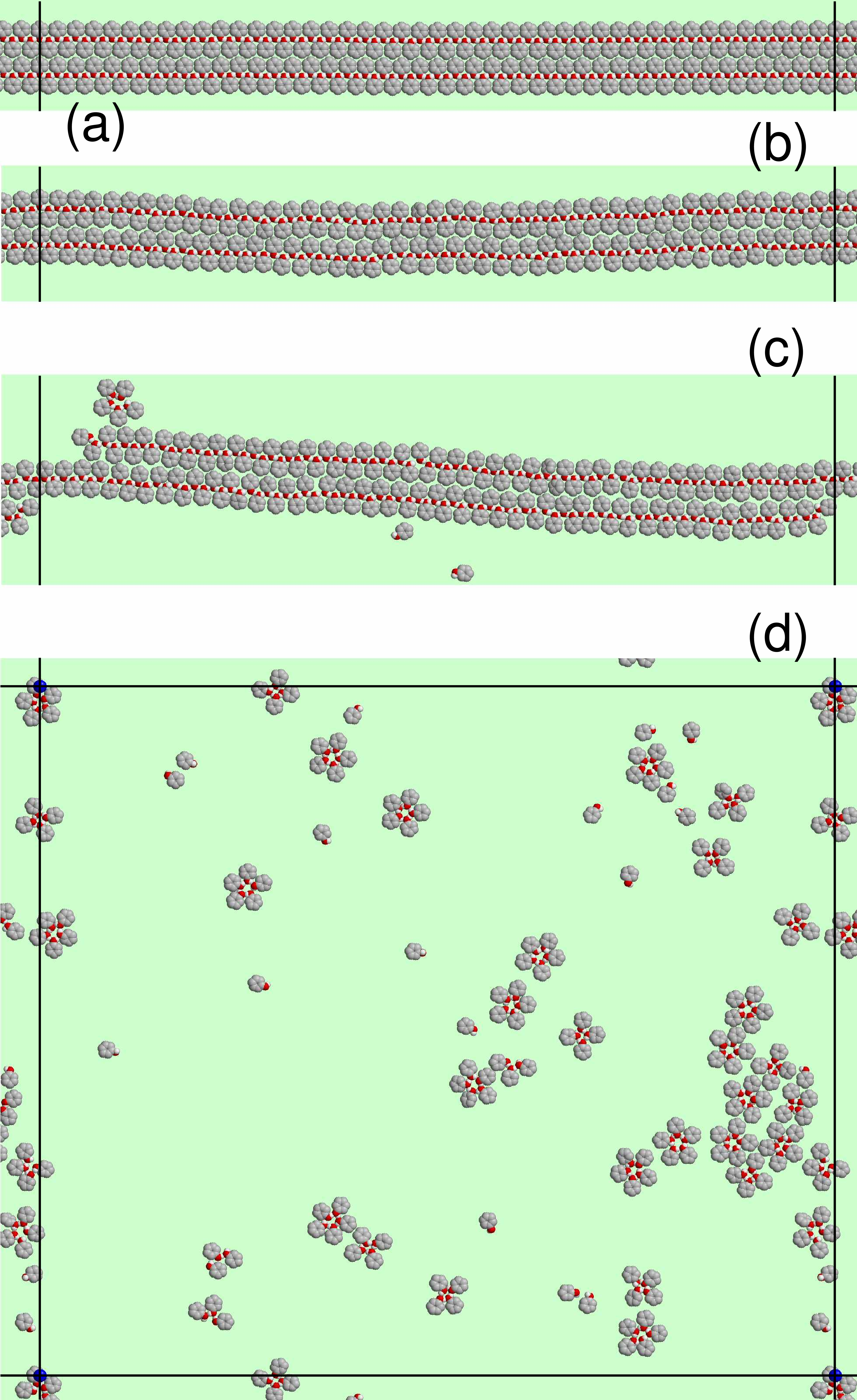}
\end{center}
\caption{\label{fig05}\protect
Structure of phenol molecules forming two adjacent hydrogen-bond chains on the flat surface of an h-BN crystal at temperatures (a) $T=10$, (b) 200, (c) 220, and (d) 240~K.
Carbon atoms are shown in gray, oxygen in red, hydrogen in white;
the flat substrate is shown in green.
Straight lines show the boundaries of the periodic cell of size $25.045\times 21.690$~nm$^2$;
the number of molecules is $N_1=176$. }
\end{figure}

The Hamiltonian of a system of $N_1$ molecules located on a flat substrate has the form
\begin{equation}
H_1=\sum_{n=1}^{N_1}\frac12 ({\bf M}\dot{\bf X}_n,\dot{\bf X_n})+E_1,
\label{f11}
\end{equation}
where the first term specifies the kinetic and the second the potential energy of the system,
\begin{equation}
E_1=\sum_{n=1}^{N_1}[V({\bf X}_n)+W({\bf X}_n)]
   +\sum_{n=1}^{N_1-1}\sum_{k=n+1}^{N_1}U({\bf X}_n,{\bf X}_k).
\label{f12}
\end{equation}
Here the vector ${\bf X}_n=\{ {\bf u}_{n,i}\}_{i=1}^{N_0}$ specifies the coordinates of the atoms of the $n$-th molecule,
${\bf M}$ is the diagonal matrix of the atomic masses of the molecule,
$V({\bf X}_n)$ and $W({\bf X}_n)$ are the deformation energy and the interaction energy with the substrate of the $n$-th molecule, and $U({\bf X}_n,{\bf X}_k)$ is the interaction energy of molecules $n$ and $k$.

The Hamiltonian of the sheet is
\begin{equation}
H_2=\sum_{n=1}^{N_2}\frac12M_0(\dot{\bf u}_n,\dot{\bf u}_n)+E_2,
\label{f13}
\end{equation}
where the first term specifies the kinetic and the second the potential energy of the sheet,
\begin{equation}
E_2=\sum_{n=1}^{N_2}[P_n+W(z_n)].
\label{f14}
\end{equation}
Here the vector ${\bf u}_n=(x_n,y_n,z_n)$ specifies the coordinate of the $n$-th united atom of the sheet.
The first term in the sum \eqref{f14} describes the interaction energy of atom $n$ with neighboring atoms of the sheet, and the second the interaction energy of the atom with the substrate.

The interaction energy of the molecules with the h-BN sheet covering them is given by the sum
\begin{equation}
E_3=\sum_{n=1}^{N_1}\sum_{i=1}^{N_0}\sum_{k=1}^{N_2}U_i(r_{nik}),
\label{f15}
\end{equation}
where the distance is $r_{nik}=|{\bf u}_{n,i}-{\bf u}_k|$, and the LJ potential
\begin{equation}
U_i(r)=\varepsilon_i[(r_i/r)^{12}-2(r_i/r)^6]
\label{f16}
\end{equation}
describes the interaction of the $i$-th atom of the molecule with a united atom of the sheet (the potential parameter values are given in Table~\ref{tab5}).
\begin{figure}[tb]
\begin{center}
\includegraphics[angle=0, width=1.0\linewidth]{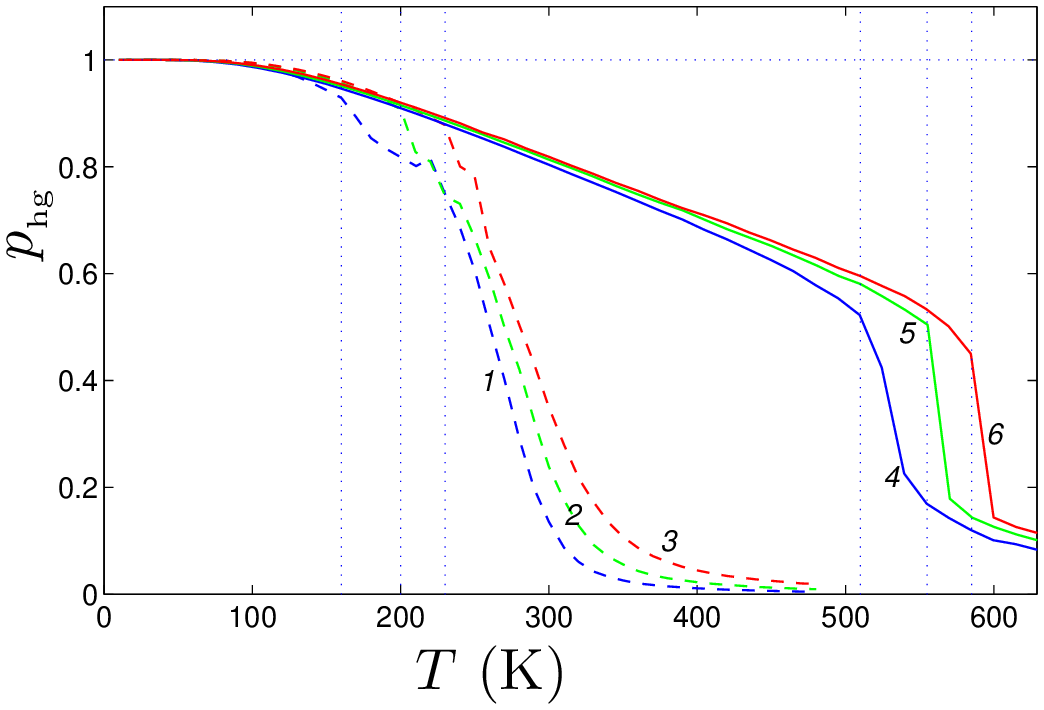}
\end{center}
\caption{\label{fig06}\protect
Temperature dependence of the specific number of hydrogen bonds $p_{\rm hg}$ for one, two, and four adjacent chains of phenol molecules (number of molecules $N_1=88$, 176, 352) lying open on a flat substrate (curves 1, 2, and 3) and covered from above by an h-BN sheet (curves 4, 5, and 6).
Vertical dashed lines show the temperature values 160, 200, 230 and 510, 555, 585~K.
}
\end{figure}

The full Hamiltonian of the two-component molecular system has the form
\begin{equation}
H=H_1+H_2+E_3,
\label{f17}
\end{equation}
and the potential energy of the system is
\begin{equation}
E=E_1+E_2+E_3.
\label{f18}
\end{equation}

To find the stationary state of a molecular system located on a flat substrate,
one must solve the problem of minimizing its potential energy
\begin{equation}
E_1\rightarrow\min: \{ {\bf u}_{n,i}\}_{n=1,i=1}^{N_1,~N_0},
\label{f19}
\end{equation}
and to find the stationary state of the encapsulated system, the problem
\begin{equation}
E\rightarrow\min: \{ {\bf u}_{n,i}\}_{n=1,i=1}^{N_1,~N_0}, \{ {\bf u}_k\}_{k=1}^{N_2}.
\label{f20}
\end{equation}

The dynamics of a thermalized open molecular system is described by the Langevin equations
\begin{eqnarray}
\label{f21}
M_{n,i}\ddot{\bf u}_{n,i}=-\frac{\partial H_1}{\partial {\bf u}_{n,i}}-\Gamma M_{n,i}\dot{\bf u}_{n,i}+\Xi_{n,i},\\
n=1,...,N_1,~~ i=1,...,N_0, \nonumber
\end{eqnarray}
where $\Gamma=1/t_r$ is the friction coefficient (relaxation time $t_r=10$~ps),
$\Xi_{n,i}=\{\xi_{n,i,k}\}_{k=1}^3$ is a three-dimensional vector of normally
distributed random forces normalized by the conditions
$$
\langle\xi_{n,i,k_1}(t_1)\xi_{l,j,k_2}(t_2)\rangle=2M_{n,i} k_BT
\Gamma\delta_{nl}\delta_{ij}\delta_{k_1k_2}\delta(t_2-t_1)
$$
($T$ is the thermostat temperature, $k_B$ is the Boltzmann constant).
\begin{figure}[tb]
\begin{center}
\includegraphics[angle=0, width=1.0\linewidth]{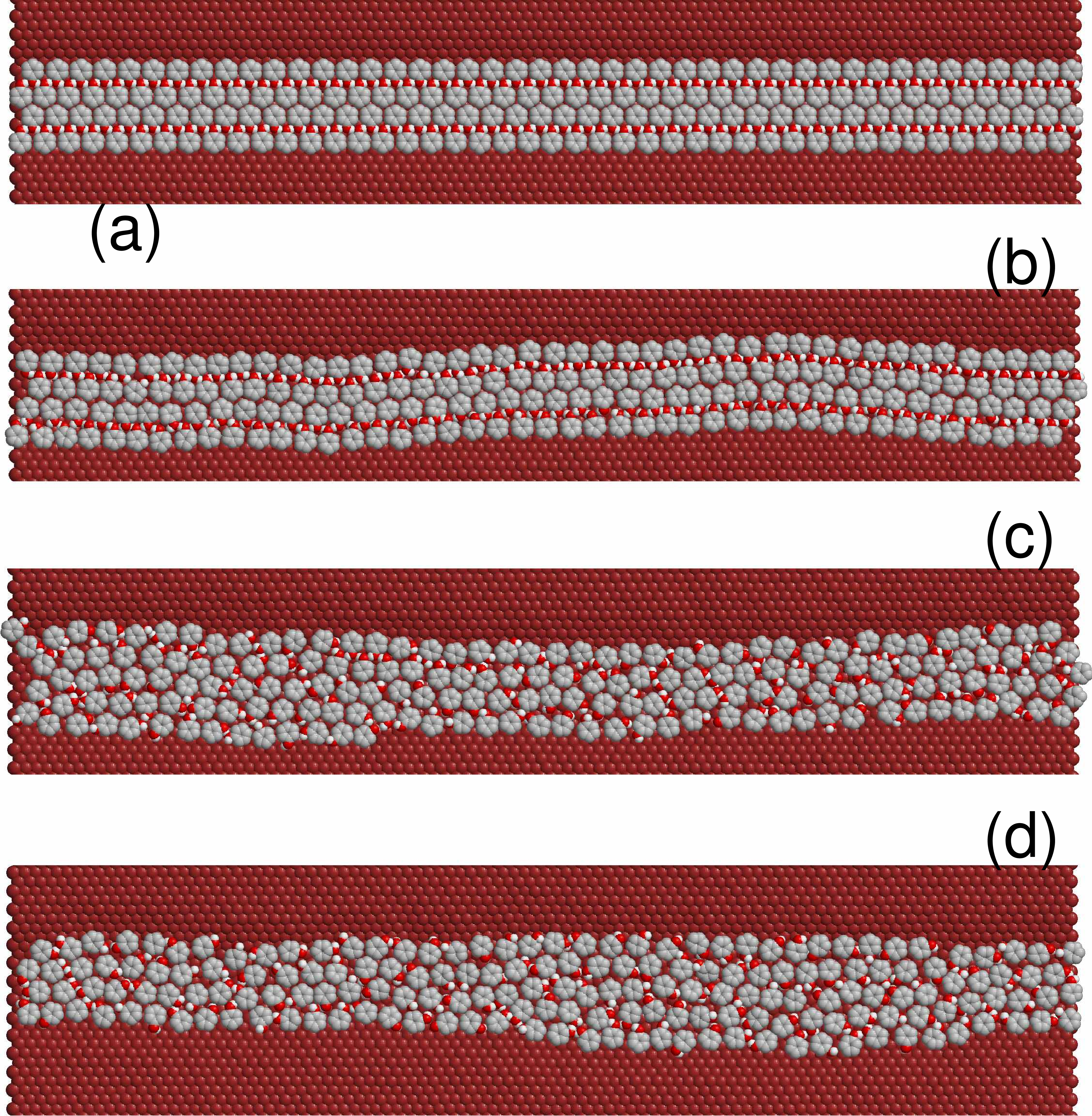}
\end{center}
\caption{\label{fig07}\protect
Structure of two adjacent chains of phenol molecules covered from above by an h-BN sheet (the sheet atoms are shown in brown) at temperatures (a) $T=15$, (b) 540, (c) 570, and (d) 600~K.
A bottom view (from the substrate side) is shown.
}
\end{figure}

The dynamics of the encapsulated molecular system is governed by the equations
\begin{eqnarray}
\label{f22}
M_{n,i}\ddot{\bf u}_{n,i}=-\frac{\partial H}{\partial {\bf u}_{n,i}}-\Gamma M_{n,i}\dot{\bf u}_{n,i}+\Xi_{n,i},\\
n=1,...,N_1,~~ i=1,...,N_0, \nonumber\\
\label{f23}
M_0\ddot{\bf u}_k=-\frac{\partial H}{\partial {\bf u}_k}-\Gamma M_0\dot{\bf u}_k+\Xi_{k},\\
k=1,...,N_2. \nonumber
\end{eqnarray}
\begin{figure}[tb]
\begin{center}
\includegraphics[angle=0, width=1.0\linewidth]{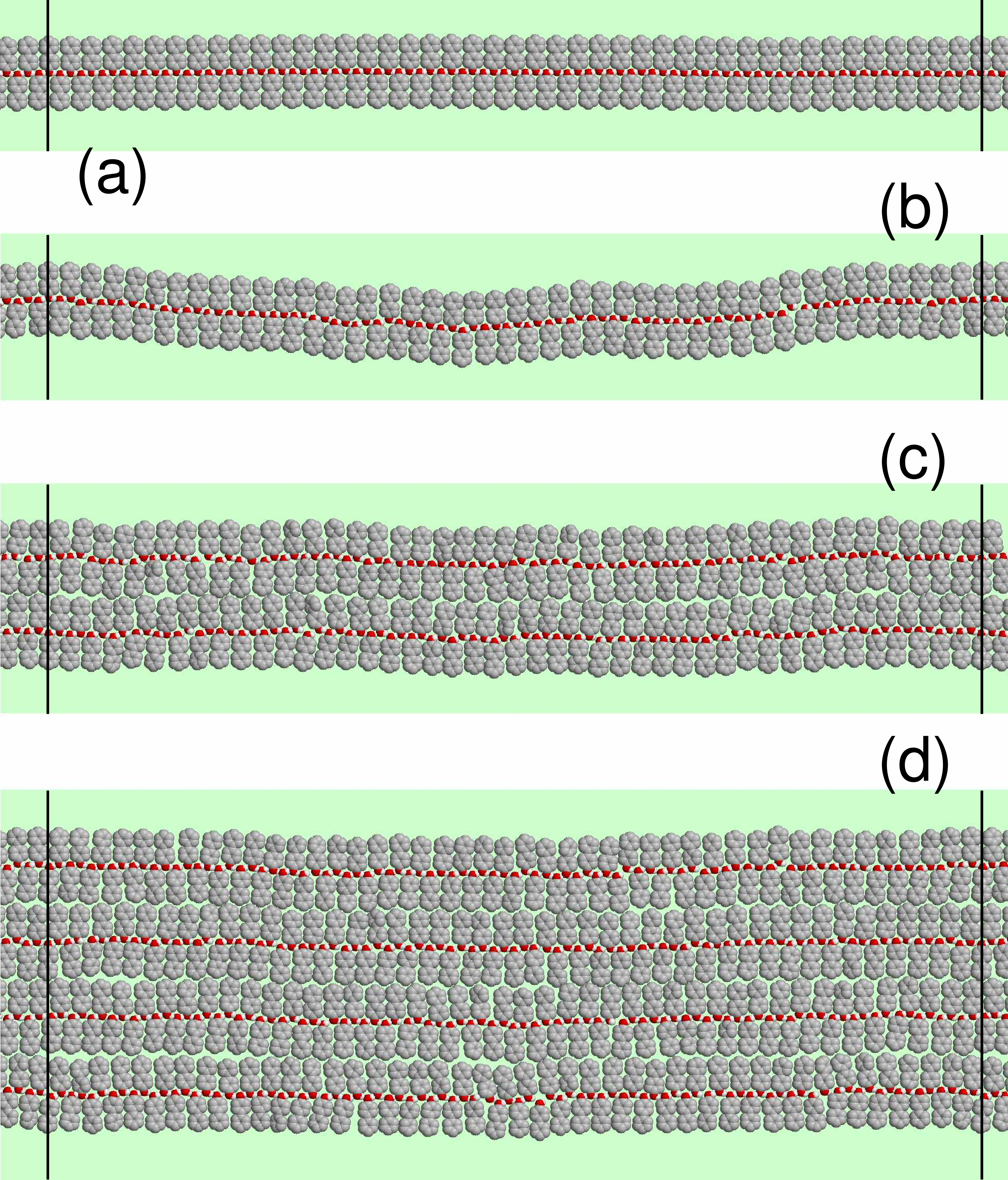}
\end{center}
\caption{\label{fig08}\protect
Structure of 4-phenylphenol molecules forming on a flat substrate one chain
at (a) temperature $T=10$ and (b) $T=210$; (c) two adjacent chains at $T=300$; (d) four chains at $T=310$~K (number of molecules $N_1=88$, 176, 352).
}
\end{figure}

The relaxation time $t_r$ specifies the intensity of energy exchange with the thermostat.
For the simulated molecular systems, the role of the thermostat is played by the flat substrate, whose coupling is governed by weak nonvalent interactions.
For such interactions, the relaxation time (energy exchange time with the thermostat) is $t_r=0.1$--$1$~ns.
For convenience of numerical simulation of the dynamics, we use a smaller value $t_r=10$~ps.
This makes it possible to significantly reduce the numerical integration time sufficient to reach the equilibrium state of the molecular system with the thermostat and to obtain reliable average values.

The solution of the energy minimization problem \eqref{f19} showed that molecules of phenol, 4-phenylphenol, paracetamol, and 4-hydroxybenzanilide on a flat substrate can form stable linear structures with hydrogen-bond chains between their hydroxyl groups.
To verify their stability against thermal fluctuations, numerical integration of the equations of motion \eqref{f21} was performed with initial conditions corresponding to the stationary states of the structures.
The stationary states of the encapsulated molecular structures were found by solving the energy minimization problems \eqref{f20}, and their stability against thermal fluctuations was verified by integrating the Langevin equations \eqref{f22}, \eqref{f23}.

The equations of motion \eqref{f21} and \eqref{f22}, \eqref{f23} were integrated numerically by the velocity Verlet method~\cite{Verlet1967} with a constant integration step $\Delta t=1$~fs.
The stationary states of the molecular systems obtained by solving the energy minimization problems \eqref{f19} and \eqref{f20} were used as initial values.
The minimization problems were solved numerically by the conjugate gradient method~\cite{Fletcher1964,Shanno1976}.
The calculations were performed on the supercomputers of the 
Supercomputer Center (JSCC) of the National Research Center ``Kurchatov Institute''.
On each supercomputer node, the integration was performed at a fixed thermostat temperature;
states for different $T$ were obtained in parallel on different nodes.
All programs were written by the author in the Fortran programming language.
The CONMIN program \cite{Shanno1976} was used only for solving the energy minimization problems \eqref{f19} and \eqref{f20}.
\begin{figure}[tb]
\begin{center}
\includegraphics[angle=0, width=1.0\linewidth]{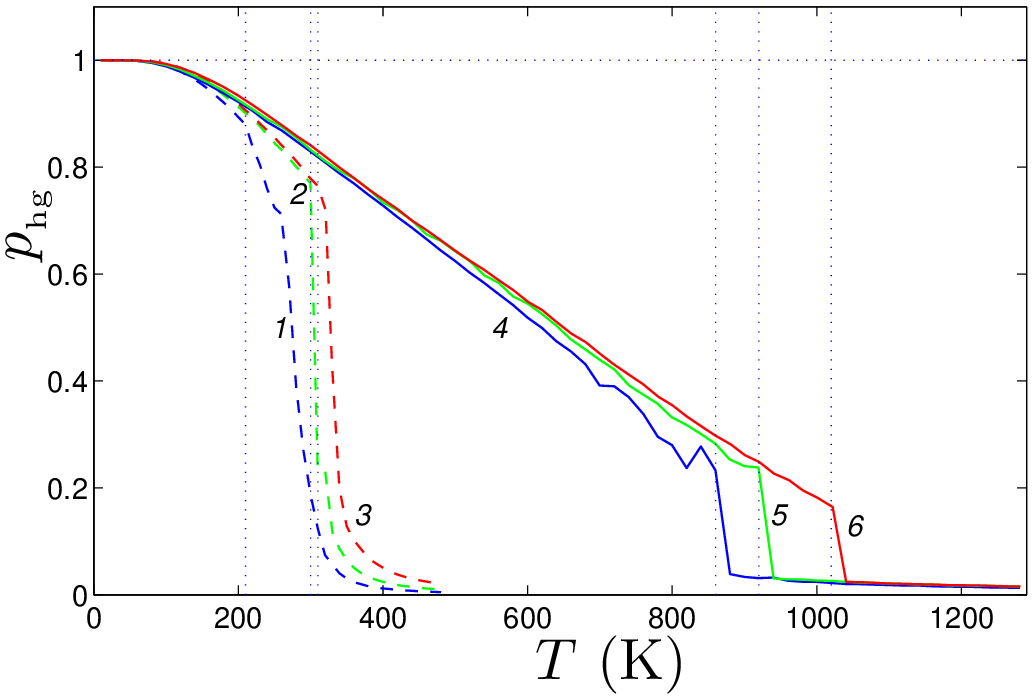}
\end{center}
\caption{\label{fig09}\protect
Temperature dependence of the specific number of hydrogen bonds $p_{\rm hg}$
for one, two, and four adjacent chains of 4-phenylphenol molecules
(number of molecules $N_1=88$, 176, 352) lying open on a flat substrate (curves 1, 2, and 3) and
covered from above by an h-BN sheet (curves 4, 5, and 6).
Vertical dashed lines show the temperature values 210, 300, 310 and 860, 920, 1020~K.
}
\end{figure}

\section{Chains of phenol molecules}

The presence of a benzene ring and a hydroxyl group in phenol molecules allows them to form stable hydrogen-bond chains on a flat substrate.
The numerical solution of the energy minimization problem \eqref{f19} showed that phenol molecules on the substrate can form stable structures of adjacent linear chains.
Each chain has in its center a zigzag hydrogen-bond chain \eqref{f1} [see Fig.~\ref{fig05}~(a)].
Each hydroxyl group participates in the formation of two hydrogen bonds: one bond is incoming, the other is outgoing.
The distance between neighboring atoms is $|{\rm OO}|=2.94$~\AA, the zigzag angle is $\angle$OOO $=150.6^\circ$, and the hydrogen bond length is $|$H$\cdots$O$|=1.97$~\AA.
The hydrogen-bond energy $E_{\rm hg}$ between two molecules (between their hydroxyl groups) can be defined as the interaction energy of the atoms with numbers $i=3$, 4, 5, 7, 8 [see Fig.~\ref{fig02} (a)].
To this end, in formula \eqref{f7} the summation should be performed only over the indices $i$ and $j$ from this set of values.
The hydrogen-bond energy between two neighboring molecules of the chain is $E_{{\rm hg},0}=-0.226$~eV.
The distance between adjacent chains (the step along the $y$ axis of the system of adjacent chains) is $h_y=11.4$~\AA.

To simulate the dynamics, we consider such structures of $N_1=88$, 176, and 352 molecules forming $N_c=1$, 2, and 4 adjacent chains (one chain contains 88 molecules).
We take a periodic computational cell of size $25.045\times 21.690$~nm$^2$ ($N_x=N_y=100$) and place
chains of molecules in it parallel to the $x$ axis (for the chosen values of $N_1$ and $N_x$, the equilibrium length of the periodic chain almost coincides with the sheet period $L_x$).

The corresponding system of equations of motion \eqref{f21} was integrated numerically over time $t=10$~ns.
After the system reached equilibrium with the thermostat, the average number of hydrogen bonds
$\bar{N}_{\rm hg}(T)$ was found (averaging was performed over one trajectory).
For definiteness, we assume that two molecules form a hydrogen bond if the interaction energy of the hydroxyl groups is $E_{\rm hg}<2E_{{\rm hg},0}/3$.

To characterize the structure of the molecular system, it is convenient to define the average number of hydrogen bonds per molecule, $p_{\rm hg}=\bar{N}_{\rm hg}(T)/N_1$.
For ideal periodically closed chains, $p_{\rm hg}=1$ (only the outgoing hydrogen bond is taken into account).
\begin{figure}[tb]
\begin{center}
\includegraphics[angle=0, width=1.0\linewidth]{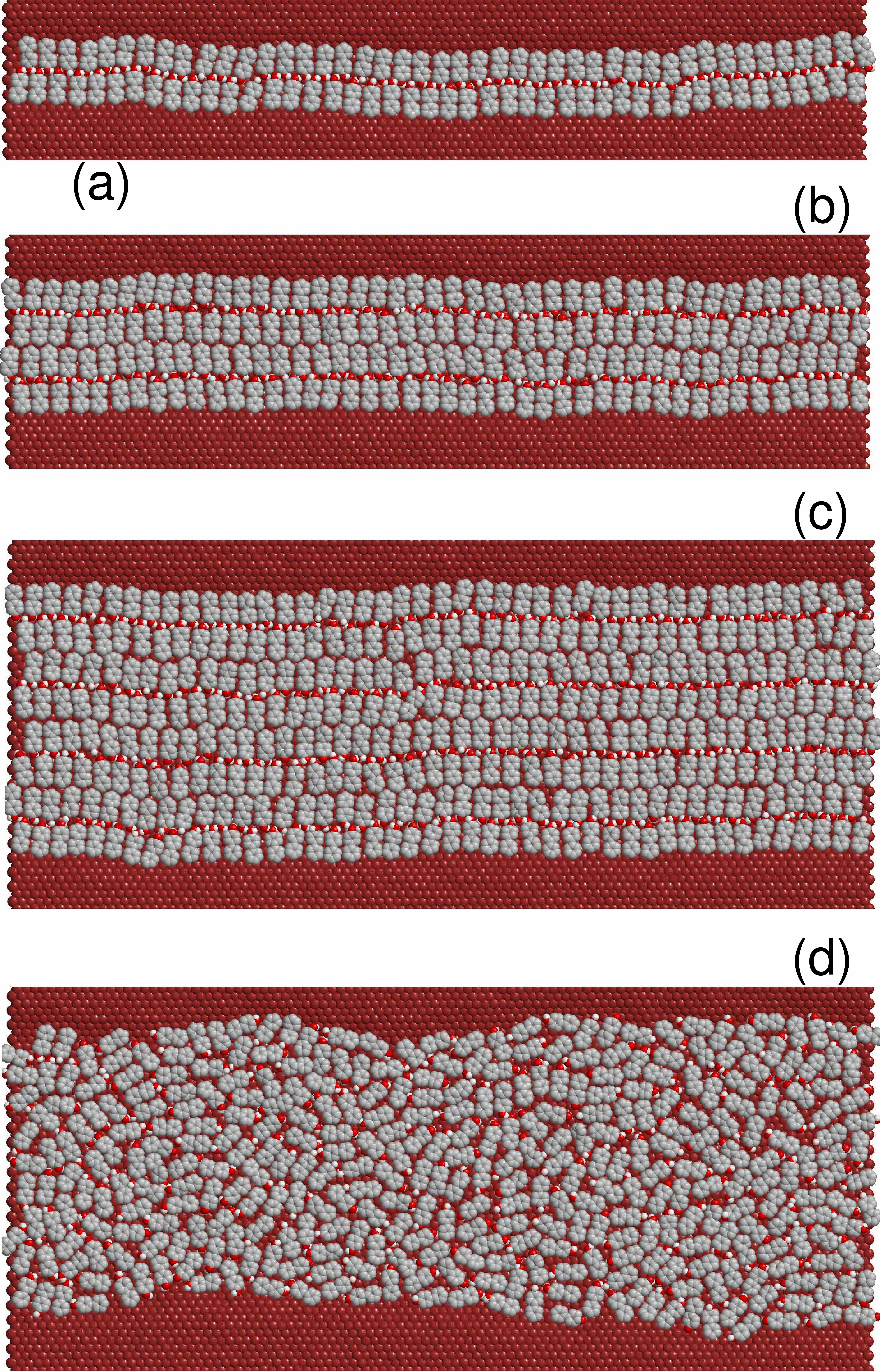}
\end{center}
\caption{\label{fig10}\protect
Structure of (a) one chain of 4-phenylphenol molecules (temperature $T=680$~K),
(b) two adjacent chains ($T=860$~K), and (c), (d) four chains ($T=920$, 1040~K)
covered from above by an h-BN sheet.
A bottom view is shown.
}
\end{figure}
\begin{figure}[tb]
\begin{center}
\includegraphics[angle=0, width=1.0\linewidth]{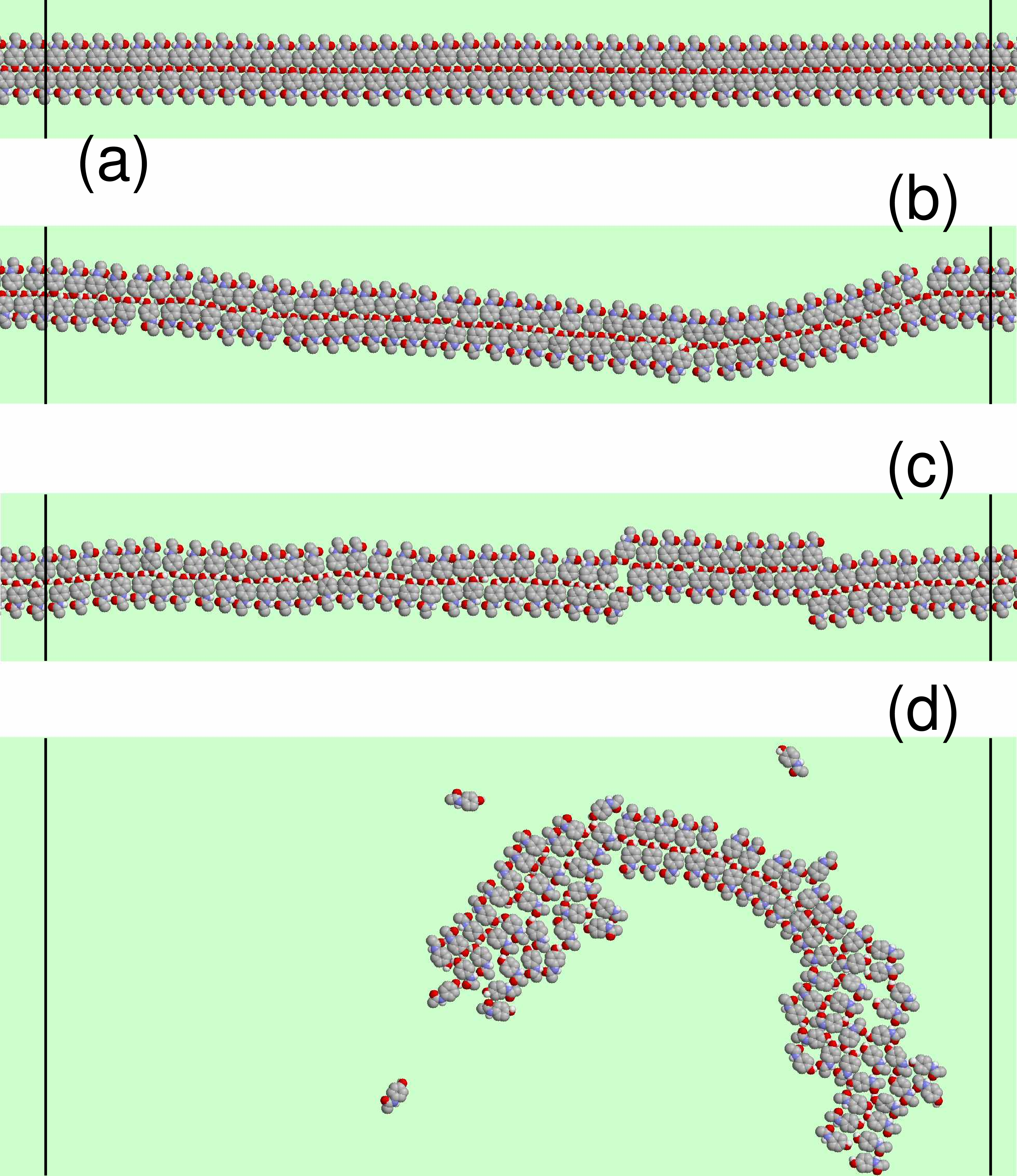}
\end{center}
\caption{\label{fig11}\protect
Structure of a single chain of paracetamol molecules (number of molecules $N_1=92$) lying on a flat substrate
at (a) temperature $T=10$, (b) 300, (c) 330, and (d) 360~K.
}
\end{figure}

The dependence of $p_{\rm hg}$ on temperature $T$ is shown in Fig.~\ref{fig06}.
As can be seen from the figure, at low temperatures $T<100$~K the specific number of hydrogen bonds is $p_{\rm hg}=1$.
Here all hydrogen bonds are retained and the molecular chains have an ideal linear form [see Fig.~\ref{fig05}~(a)].
As the temperature increases, some hydrogen bonds can temporarily weaken, and the chains can bend [Fig.~\ref{fig05}~(b)].
When a threshold temperature $T_o$ is reached, chain disruption occurs:
some hydrogen bonds break irreversibly and the chains disintegrate into unconnected fragments [Fig.~\ref{fig05}~(c,d)].
At this temperature, a sharp decrease in the number of hydrogen bonds begins.
For phenol molecules, the disruption of one open hydrogen-bond chain begins at temperature $T_{o,1}=160$, of two adjacent chains at $T_{o,2}=200$, and of four at $T_{o,4}=230$~K.

Let us cover the chains of phenol molecules lying on the flat substrate with a rectangular h-BN sheet of size $25.045\times 26.028$~nm$^2$ ($N_x=100$, $N_y=120$, number of united atoms of the sheet $N_2=N_xN_y=12000$).
Numerical integration of the equations of motion \eqref{f22}, \eqref{f23} showed that the encapsulated
chains become significantly more stable.
The interaction of the molecules with the sheet covering them creates a high internal pressure that prevents the chains from disintegrating even at $T>600$~K.
In contrast to open chains, which disintegrate into fragments, encapsulated chains
retain their linear structure, but at $T>T_c$ only their melting occurs, accompanied by a sharp decrease in the number of hydrogen bonds --- see Fig.~\ref{fig07}.
As can be seen from Fig.~\ref{fig06}, the melting of hydrogen-bond chains for one chain of molecules begins at temperature $T_{c,1}=510$, for two adjacent chains at $T_{c,2}=555$, and for four at $T_{c,4}=585$~K.
\begin{figure}[tb]
\begin{center}
\includegraphics[angle=0, width=1.0\linewidth]{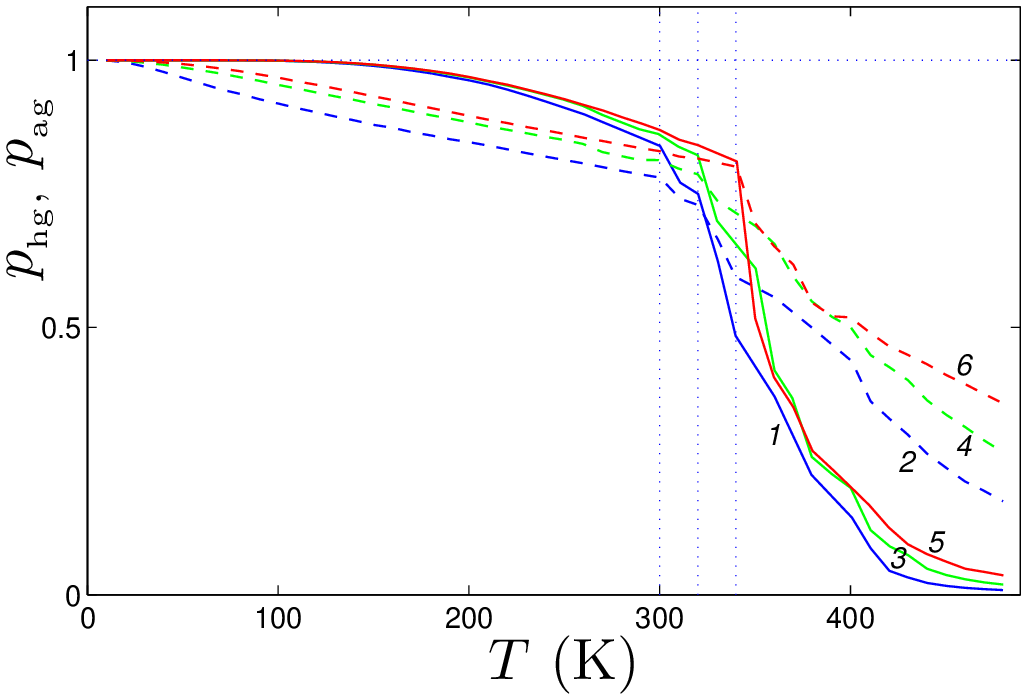}
\end{center}
\caption{\label{fig12}\protect
Temperature dependence of the specific numbers of hydrogen bonds $p_{\rm hg}$ and $p_{\rm ag}$
(solid curves 1, 3, 5 and dashed curves 2, 4, 6) for one, two, and four adjacent chains of paracetamol molecules (number of molecules $N_1=92$, 184, 368) lying open on a flat substrate (curves 1, 2; 3, 4; and 5, 6).
Vertical lines show the temperature values $T=300$, 320, and 340~K.
}
\end{figure}

\section{Chains of 4-phenylphenol molecules}

The stability of molecular chains located on flat substrates can be increased by enhancing their interaction with the substrate.
To this end, for example, another benzene ring can be added to the phenol molecule.
However, if a new ring is attached directly to the face of the existing one, forming a single planar aromatic system, its dimensions will not allow long continuous hydrogen-bond chains to form on a flat substrate --- steric effects will hinder their formation.
An increase in the transverse size of the molecule can be avoided if the benzene ring
(the phenyl group --\ce{C6H5}) is added through a single valence bond \ce{C-C}, forming the 4-phenylphenol (4-PP) molecule --- see Fig.~\ref{fig02} (b).
\begin{figure}[tb]
\begin{center}
\includegraphics[angle=0, width=1.0\linewidth]{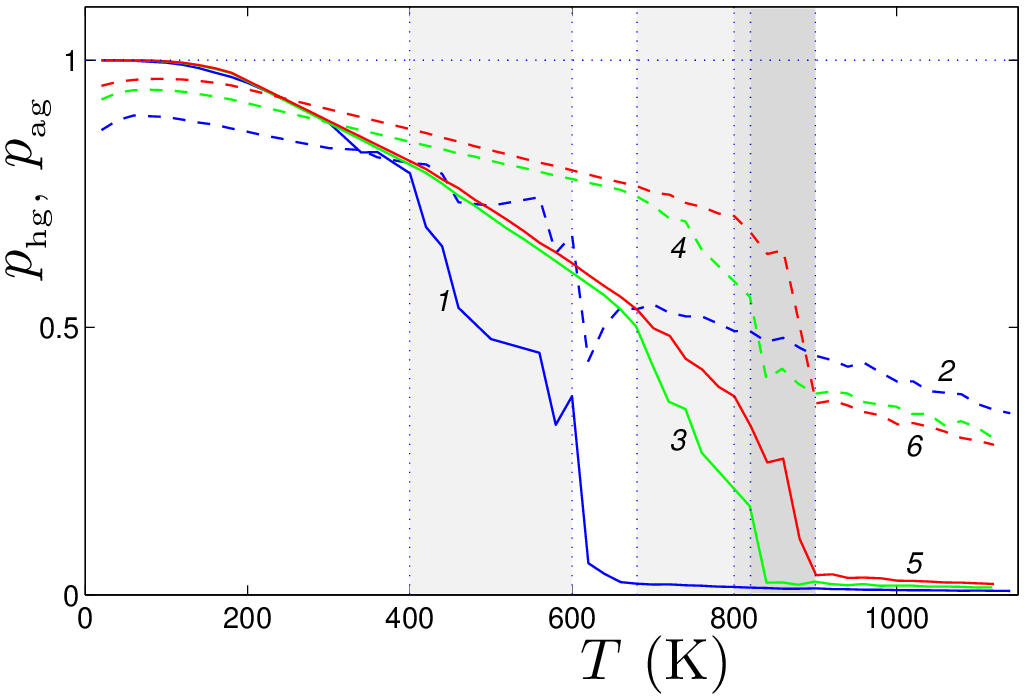}
\end{center}
\caption{\label{fig13}\protect
Temperature dependence of the specific numbers of hydrogen bonds $p_{\rm hg}$ and $p_{\rm ag}$ (solid curves 1, 3, 5 and dashed curves 2, 4, 6) for one, two, and four adjacent chains of paracetamol molecules (number of molecules $N_1=92$, 184, 368) lying on a flat substrate and covered from above by an h-BN sheet (curves 1, 2; 3, 4; and 5, 6).
Gray bands show the temperature intervals $T\in[T_1,T_2]$ in which continuous melting of the molecular chains occurs: for one chain [400,600]; for two [680,820] and [800,900]~K for four chains.
}
\end{figure}

The numerical solution of the energy minimization problem \eqref{f19} showed that 4-PP molecules, like the phenol molecule, can form structures of adjacent parallel molecular chains containing in their center hydrogen-bond chains \eqref{f1} --- see Fig.~\ref{fig08}.
Here the bond chain has a zigzag form with distance $|$OO$|=2.95$~\AA~ and angle $\angle$OOO $=151^\circ$.
The energy of one hydrogen bond is $E_{{\rm hg},0}=-0.224$~eV.
The distance between adjacent chains is $h_y=19.95$~\AA.

To simulate the dynamics, we consider such structures of $N_1=88$, 176, and 352 4-PP molecules forming $N_c=1$, 2, and 4 adjacent chains.
We take a periodic computational cell of size $25.295\times 26.028$~nm$^2$ ($N_x=101$, $N_y=120$) and place chains of molecules in it parallel to the $x$ axis.

The dependence of the specific number of hydrogen bonds $p_{\rm hg}$ on temperature is shown in Fig.~\ref{fig09}.
As can be seen from the figure, the destruction of one open chain begins at temperature $T_{o,1}=210$,
of two adjacent chains at $T_{o,2}=300$, and of four at $T_{o,4}=310$~K.
The appearance of the molecular chains at these temperatures is shown in Fig.~\ref{fig08}~(b,c).

Let us cover the chains from above with an h-BN sheet of size $25.295\times 26.028$~nm$^2$ (number of united atoms of the sheet $N_2=N_xN_y=12120$).
Simulation of the dynamics of the two-component molecular system showed that the encapsulated chains
retain their linear form in the considered temperature range.
Only their melting can occur, with disorientation of neighboring molecules and breaking of most hydrogen bonds [see Fig.~\ref{fig10}~(c,d)].
As can be seen from Fig.~\ref{fig09}, the melting of the hydrogen-bond chain for one encapsulated chain is completed at temperature $T_{c,1}=860$, and for two and four adjacent chains at $T_{c,2}=920$ and $T_{c,4}=1020$~K.
\begin{figure}[tb]
\begin{center}
\includegraphics[angle=0, width=1.0\linewidth]{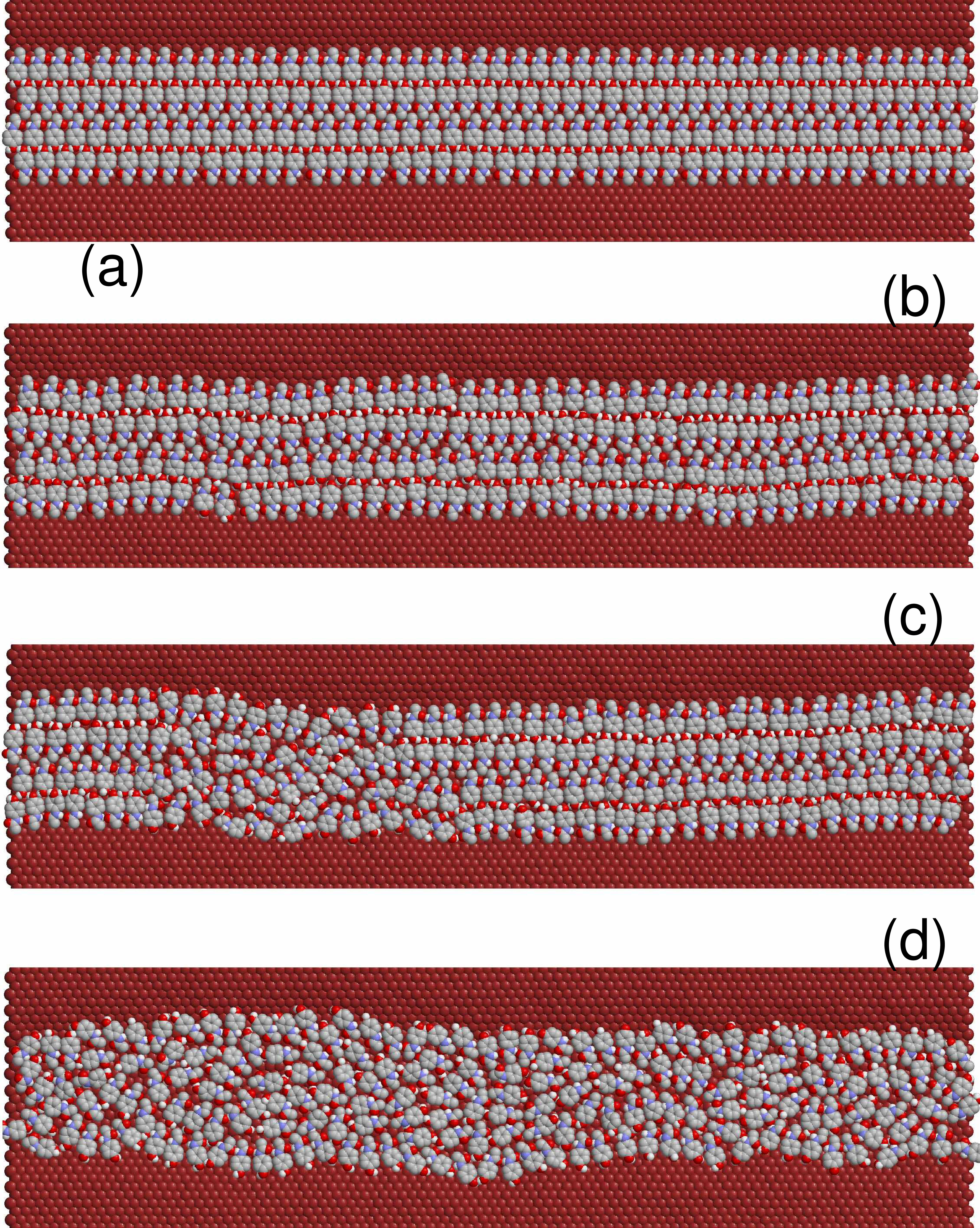}
\end{center}
\caption{\label{fig14}\protect
Structure of two adjacent chains of paracetamol molecules (number of molecules $N_1=184$)
covered from above by an h-BN sheet at (a) $T=20$, (b) 680, (c) 720, and (d) 840~K.
A bottom view is shown.
}
\end{figure}
\begin{figure}[tb]
\begin{center}
\includegraphics[angle=0, width=1.0\linewidth]{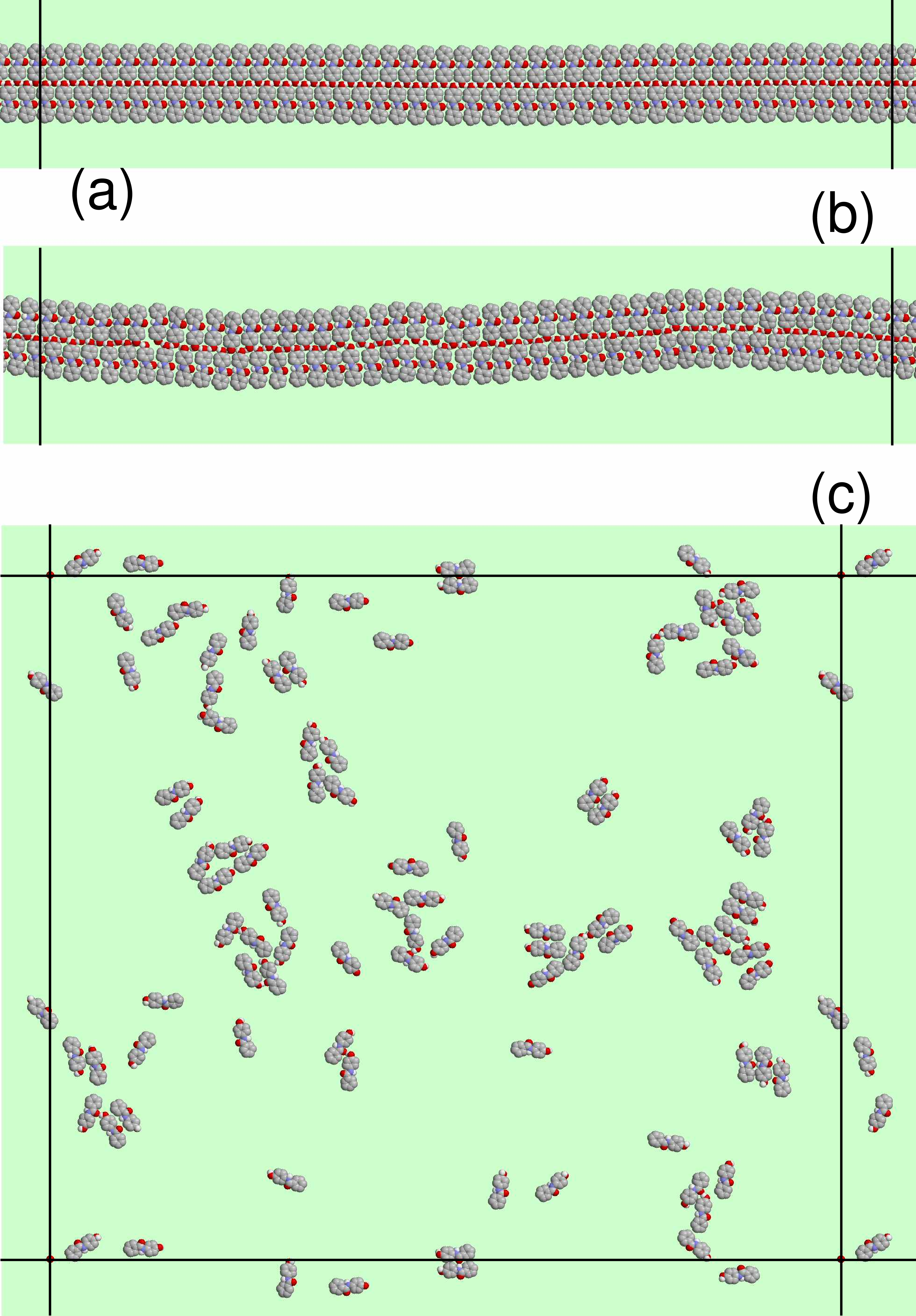}
\end{center}
\caption{\label{fig15}\protect
Structure of a single chain of 4-HBA molecules (number of molecules $N_1=92$)
lying on a flat substrate at (a) temperature $T=10$, (b) 390, and (c) 430~K.
Straight lines show the boundaries of the periodic cell of size $26.045\times 22.557$~nm$^2$.
}
\end{figure}

\section{Chains of paracetamol molecules}

The numerical solution of the potential energy minimization problem \eqref{f19} showed that paracetamol molecules on a flat substrate can form linear chains in which each molecule participates in the formation of four hydrogen bonds (two incoming and two outgoing).
Here one hydrogen-bond chain \eqref{f1} is formed by hydroxyl groups OH (hg), and the second chain
by amide groups of the molecules OCNH (ag).
As a result, one chain of molecules consists of three hydrogen-bond chains: two chains
\begin{equation}
\ce{H-N-C=O}\cdots\ce{H-N-C=O}\cdots\ce{H-N-C=O}\cdots
\label{f24}
\end{equation}
and the chain \eqref{f1} located between them [see Fig.~\ref{fig11}~(a,b)].

The hydrogen-bond energy between hydroxyl groups $E_{\rm hg}$ can be defined as the interaction energy of
the atoms with numbers $i=3$, 4, 5, 12, and 13, and the bond energy between amide groups $E_{\rm ag}$ as the interaction energy of the atoms with $i=1$, 7, 8, 9, 10, and 11 [see Fig.~\ref{fig02} (c)].
The hydrogen-bond chain of hydroxyl groups has a zigzag form with distance $|$OO$|=2.88$~\AA~ and zigzag angle $\angle$OOO $=146^\circ$.
The bond length is $|$O$\cdots$H$|=1.90$~\AA, and the energy is $E_{{\rm hg},0}=-0.226$~eV.
The hydrogen-bond chain of amide groups \eqref{f24} has a linear form, with chain step $|$OO$|=5.55$~\AA.
The hydrogen bond length between amide groups is $|$O$\cdots$H$|=2.35$~\AA, and the bond energy is $E_{{\rm ag},0}=-0.194$~eV.
The distance between adjacent molecular chains is $h_y=18.5$~\AA.

To simulate the dynamics, we consider such structures of $N_1=92$, 184, and 368 paracetamol molecules forming $N_c=1$, 2, and 4 adjacent chains (one chain consists of 92 molecules).
We take a periodic computational cell of size $25.546\times 22.123$~nm$^2$ ($N_x=102$, $N_y=102$) and place chains of molecules in it parallel to the $x$ axis.
For each temperature value, we find the average number of hydrogen bonds between hydroxyl $\bar{N}_{\rm hg}(T)$ and amide $\bar{N}_{\rm ag}(T)$ groups.
For definiteness, we assume that two molecules form a hydrogen bond between hydroxyl groups
if the interaction energy is $E_{\rm hg}<2E_{{\rm hg},0}/3$, and a hydrogen bond between amide groups if $E_{\rm ag}<2E_{{\rm ag},0}/3$.

To characterize the structure of the molecular system, it is convenient to define the average number of each type of hydrogen bond per molecule: $p_{\rm hg}=\bar{N}_{\rm hg}(T)/N_1$
and $p_{\rm ag}=\bar{N}_{\rm ag}(T)/N_1$.
For ideal periodically closed chains, the specific numbers of bonds are $p_{\rm hg}=1$, $p_{\rm ag}=1$
(only outgoing bonds are taken into account: one hydrogen bond per hydroxyl group and one per amide group of the molecule).
\begin{figure}[tb]
\begin{center}
\includegraphics[angle=0, width=1.0\linewidth]{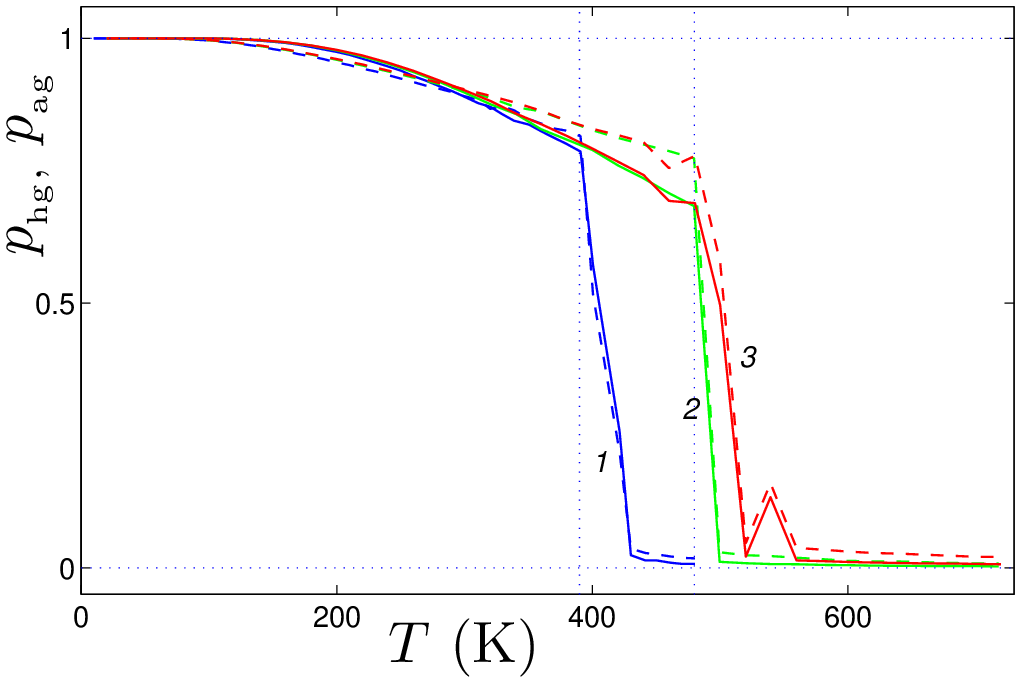}
\end{center}
\caption{\label{fig16}\protect
Temperature dependence of the specific numbers of hydrogen bonds $p_{\rm hg}$ and $p_{\rm ag}$ (solid and dashed curves) for one, two, and four adjacent chains of 4-HBA molecules (number of molecules $N_1=92$, 184, 368) lying open on a flat substrate (curves 1, 2, and 3).
Vertical lines show the temperature values 390 and 480~K.
}
\end{figure}

The dependences $p_{\rm hg}(T)$ and $p_{\rm ag}(T)$ obtained from the simulations of the dynamics of molecular chains are shown in Figs.~\ref{fig12} and \ref{fig13}.
As can be seen from Fig.~\ref{fig12}, for chains lying open on a flat substrate at low temperatures the specific numbers of hydrogen bonds are $p_{\rm hg}=1$, $p_{\rm ag}=1$.
Here all hydrogen bonds between the hydroxyl and amide groups of the molecules are retained, and the chains themselves have an ideal linear form [see Fig.~\ref{fig11}~(a)].
As the temperature increases, some hydrogen bonds can temporarily weaken and the chains can bend [Fig.~\ref{fig11}~(b)].
When a threshold temperature is reached, breaking of hydrogen bonds begins, leading to the disintegration of the chains into unconnected fragments [Fig.~\ref{fig11}~(c,d)].
At this temperature, a sharp decrease in the number of hydrogen bonds begins.
As can be seen from Fig.~\ref{fig12}, for paracetamol molecules lying open on a flat substrate, the disruption of one chain begins at temperature $T_{o,1}=300$, of two adjacent chains at $T_{o,2}=320$, and of four at $T_{o,4}=340$~K.
\begin{figure}[tb]
\begin{center}
\includegraphics[angle=0, width=1.0\linewidth]{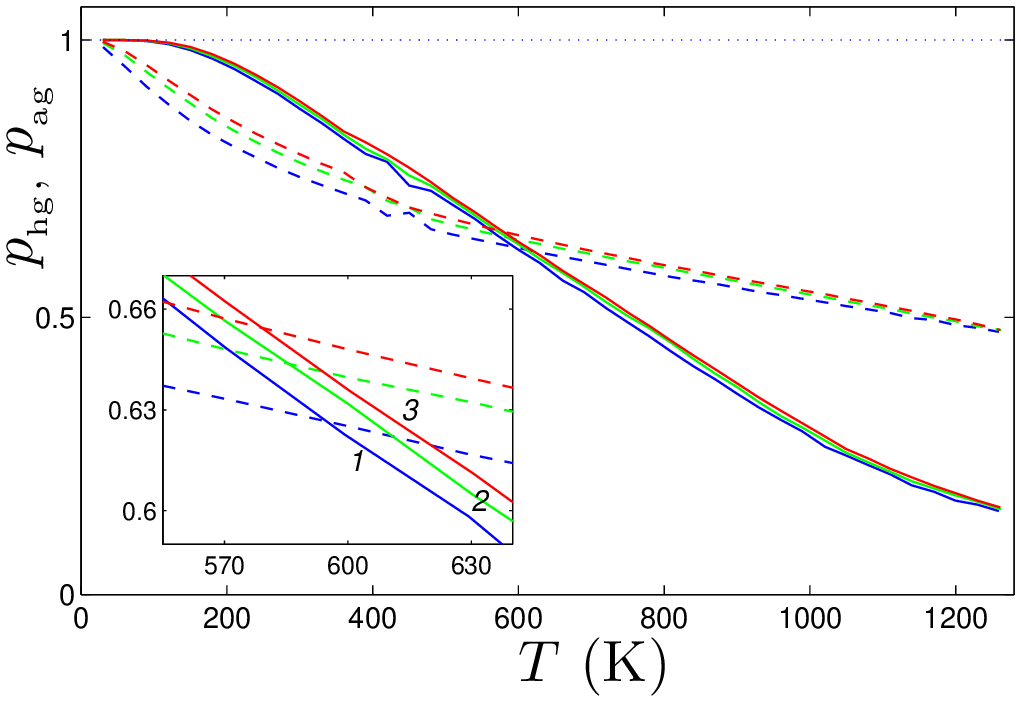}
\end{center}
\caption{\label{fig17}\protect
Temperature dependence of the specific numbers of hydrogen bonds $p_{\rm hg}$ and $p_{\rm ag}$ (solid and dashed curves) for one, two, and four adjacent chains of 4-HBA molecules (number of molecules $N_1=92$, 184, 368) covered by an h-BN sheet (curves 1, 2, and 3).
}
\end{figure}

Let us cover the system of paracetamol molecular chains lying on the flat substrate with a rectangular h-BN sheet of size
$25.546\times 26.028$~nm$^2$ ($N_x=102$, $N_y=120$).
Numerical simulation of the dynamics of the two-component molecular system showed that the encapsulated chains become significantly more stable. As can be seen from Fig.~\ref{fig13}, the melting of one encapsulated chain begins at temperature $T_{m,1}=400$, the melting of two adjacent chains at 680~K, and of four at 800~K. At $T<T_{m,1}$, the hydrogen-bond chains remain stable against thermal fluctuations, see Fig.~\ref{fig14}~(a,b).
The melting of the system of parallel chains occurs continuously in the temperature interval $[T_{m,1},T_{m,2}]$
(the temperature $T_{m,2}=600$, 820, and 900~K for one, two, and four chains, respectively).
Here the molecular system splits into two parts: one part retains the structure with linear hydrogen-bond chains, while the other passes into a disordered state with broken hydrogen bonds [see Fig.~\ref{fig14}~(c)]. As the temperature increases, the fraction of the molten (disordered) part of the system increases.
At $T=T_{m,2}$, the entire molecular system passes into the molten state --- see Fig.~\ref{fig14}~(d). At the same time, at all considered temperatures, the encapsulated chain retains its linear single-layer (planar) form.

Note that such a melting scenario is characteristic of 2D crystals.
While in three-dimensional space the melting of a crystal is a first-order phase transition and occurs at a definite temperature $T_m$, the melting of a two-dimensional system occurs continuously in a certain temperature interval $[T_{m,1},T_{m,2}]$ \cite{Mak2006,Ryzhov2017,Tsiok2020,Toledano2021,Zhang2021}.
\begin{figure}[tb]
\begin{center}
\includegraphics[angle=0, width=1.0\linewidth]{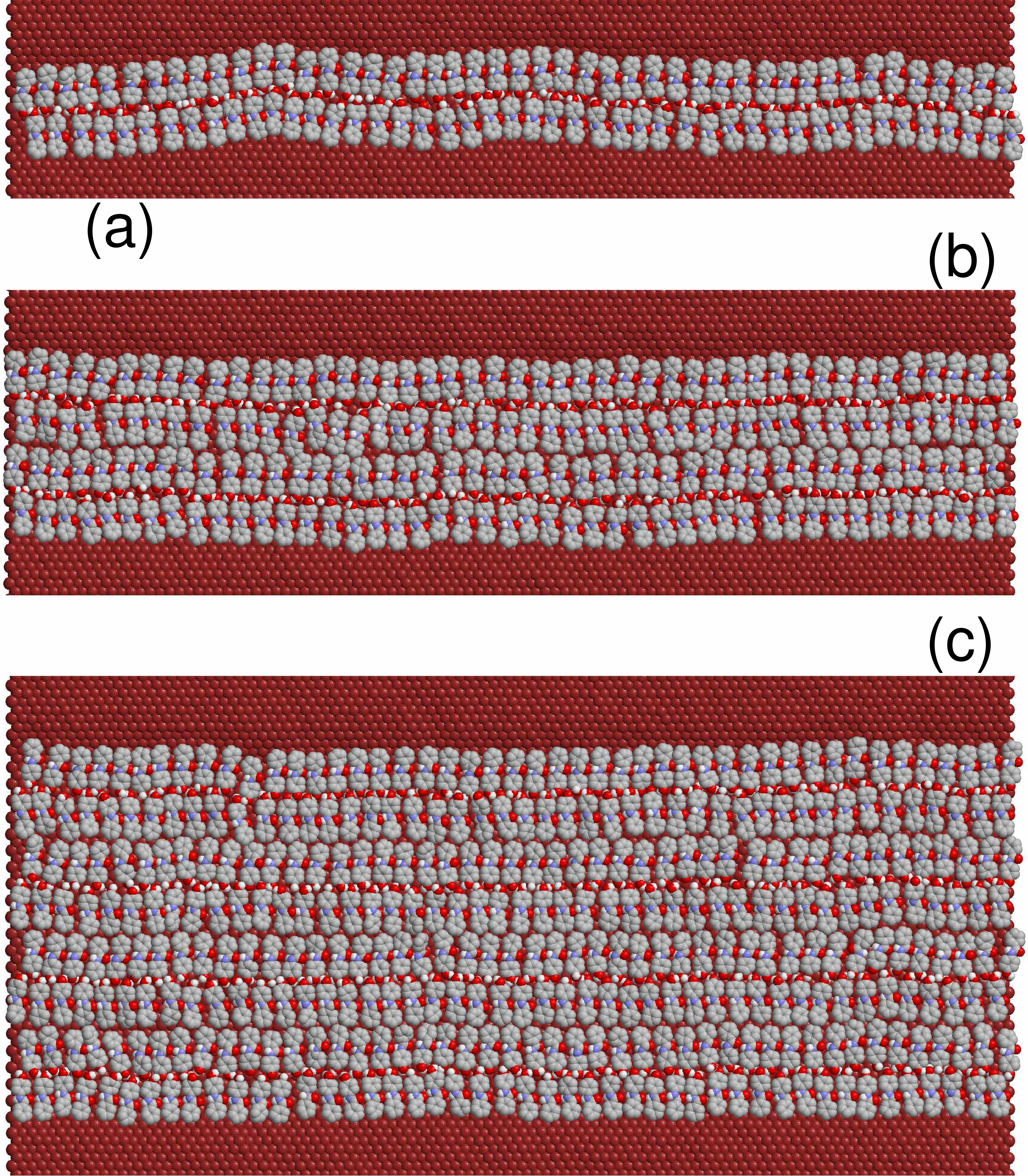}
\end{center}
\caption{\label{fig18}\protect
Structure of encapsulated (a) one, (b) two, and (c) four adjacent chains of
4-HBA molecules (number of molecules $N_1=92$, 184, 368) at temperature $T=1260$~K.
}
\end{figure}

\section{Chains of 4-hydroxybenzanilide molecules}

The thermal stability of paracetamol molecular chains located on a flat substrate can be further increased by enhancing their interaction with the substrate.
To this end, it is sufficient to replace the methyl group --\ce{CH3} in each molecule with a phenyl group --\ce{C6H5}, thereby obtaining molecules of 4-hydroxybenzanilide (4-HBA) with two benzene rings [see Fig.~\ref{fig02}~(d)].

The numerical solution of the energy minimization problem \eqref{f19} showed that 4-HBA molecules, like paracetamol molecules, can form linear chains with three hydrogen-bond chains: two chains of amide groups \eqref{f24} and the hydroxyl-group chain \eqref{f1} located between them [see Fig.~\ref{fig15}~(a)].
Here the hydrogen-bond chain \eqref{f1} has a zigzag form with distance $|$OO$|=2.92$~\AA \ and angle $\angle$OOO $=149^\circ$.
The hydrogen-bond energy between hydroxyl groups is $E_{{\rm hg},0}=-0.228$~eV, and the bond length is $|$O$\cdots$H$|=1.94$~\AA.
The hydrogen-bond chain \eqref{f24} has a linear form, with chain step $|$OO$|=5.66$~\AA.
The hydrogen-bond energy between amide groups is $E_{{\rm ag},0}=-0.178$~eV, and the bond length is $|$O$\cdots$H$|=2.47$~\AA. The distance between adjacent chains is $h_y=24.53$~\AA.

To simulate the dynamics, we consider such structures of $N_1=92$, 184, and 368 molecules forming $N_c=1$, 2, and 4 adjacent chains.
We take a periodic computational cell of size $26.047\times 22.557$~nm$^2$ and place chains of molecules in it parallel to the $x$ axis.

To simulate the dynamics of the molecular system with open chains, we numerically integrate the equations of motion \eqref{f21} at various temperature values.
The resulting temperature dependence of the specific numbers of hydrogen bonds $p_{\rm hg}$, $p_{\rm ag}$ is shown in Fig.~\ref{fig16}.
As can be seen from Fig.~\ref{fig16}, the disruption of one open chain begins at temperature $T_{o,1}=390$,
and of two and four adjacent chains at $T_{o,2}=T_{o,4}=480$~K.
The appearance of the structure of a single chain before and after the onset of disruption is shown in Fig.~\ref{fig15}~(b,c).

Let us cover the system of chains lying on the flat substrate with a rectangular h-BN sheet of size $26.047\times 27.763$~nm$^2$
($N_x=104$, $N_y=128$).
Simulation of the dynamics of the two-component molecular system showed that encapsulated chains of 4-HBA molecules possess the highest thermal stability.
Their hydrogen-bond chains are retained at all considered temperatures $T\le 1260$~K
(see Figs.~\ref{fig17} and \ref{fig18}).
Here melting of the encapsulated chains does not occur, and all hydrogen-bond chains are retained (only
temporary breaking of some hydrogen bonds is possible).

\section{Conclusion}

The performed numerical simulation of the dynamics of linear molecular chains located on a flat substrate (on the surface of an h-BN crystal) shows that molecules containing benzene rings, amide and hydroxyl groups in their structure can form stable hydrogen-bond chains of hydroxyl groups \eqref{f1}.
Such chains can be formed by molecules of phenol (PhOH), 4-phenylphenol (4-PP), paracetamol (PCM), and 4-hydroxybenzanilide (4-HBA).
The benzene rings of these molecules, on the one hand, ensure their strong interaction with the flat substrate and, on the other hand, do not hinder the formation of long hydrogen-bond chains.
The disruption of such chains lying open on a flat substrate occurs at temperatures above $T_1=160$, 210, 300, and 390~K for PhOH, 4-PP, PCM, and 4-HBA molecules, respectively.
Covering the chains with an h-BN sheet (their encapsulation) significantly enhances their thermal stability.
Such encapsulated molecular structures retain their hydrogen-bond chains up to temperatures $T_2=510$, 860, and 600~K for PhOH, 4-PP, and PCM molecules, while the hydrogen-bond chains of 4-HBA molecules remain stable at all considered $T\le 1260$~K.
Thus, hydrogen-bond chains can be retained at high temperatures.
Note that the melting point of phosphoric acid, most frequently used in proton-exchange membranes (PEMs), is only 315~K.
Hydrogen-bond chains \eqref{f1} are efficient pathways for proton transfer.
Therefore, the considered molecular chains encapsulated by h-BN sheets can be used to create anhydrous proton-exchange membranes possessing high thermal stability.
The most promising are chains of 4-hydroxybenzanilide molecules, since their hydrogen-bond chains possess the highest thermal stability.

To date, PEMs capable of operating at a temperature of 523~K have been created~\cite{He2026,Stepanov2026}.
In these membranes, proton transport is carried out along hydrogen-bond chains formed by phosphoric acid molecules.
The performed simulations allow us to conclude that multilayer structures of h-BN sheets and paracetamol and 4-hydroxybenzanilide molecules can be used to create new PEMs capable of operating at higher temperatures.
\\ \\

{\bf Acknowledgements}\\

Computational facilities were provided by the Joint Supercomputer Center (JSCC) of the National Research Center ``Kurchatov Institute''. 
The research was funded by the Russian Science Foundation (RSF) (project No. 25-73-20038).

\end{document}